\documentclass{aa} 

\usepackage{graphicx}
\usepackage{multirow}
\usepackage{txfonts}
\usepackage[dvipsnames]{xcolor}
\usepackage{caption,subcaption} 

\usepackage{float}

\begin{document}

   \title{Characterisation of the hydrospheres of TRAPPIST-1 planets}


   \author{Lorena Acuña\inst{1}
          \and Magali Deleuil \inst{1}
          \and Olivier Mousis \inst{1}
          \and Emmanuel Marcq \inst{2}
          \and Maëva Levesque \inst{1}
          \and Artyom Aguichine \inst{1}
           }

   \institute{Aix Marseille Univ, CNRS, CNES, LAM, Marseille, France\\
              \email{lorena.acuna@lam.fr}
         \and
             LATMOS/IPSL, UVSQ Université Paris-Saclay, Sorbonne Université, CNRS, Guyancourt, France
}

   \date{Received 12 November 2020; accepted -}

 
  \abstract
   {Planetary mass and radius data are showing a wide variety in densities of low-mass exoplanets. This includes sub-Neptunes, whose low densities can be explained with the presence of a volatile-rich layer. Water is one of the most abundant volatiles, which can be in the form of different phases depending on the planetary surface conditions. To constrain their composition and interior structure, it is required to develop models that calculate accurately the properties of water at its different phases.}   
   {We present an interior structure model that includes a multiphase water layer with steam, supercritical and condensed phases. We derive the constraints for planetary compositional parameters and their uncertainties, focusing on the multiplanetary system TRAPPIST-1, which presents both warm and temperate planets.
}
   {We use a 1D steam atmosphere in radiative-convective equilibrium with an interior whose water layer is in supercritical phase self-consistently.  For temperate surface conditions, we implement liquid and ice Ih to ice VII phases in the hydrosphere. We adopt a MCMC inversion scheme to derive the probability distributions of core and water compositional parameters
}
   {We refine the composition of all planets and derive atmospheric parameters for planets b and c. The latter would be in a post-runaway greenhouse state and could be extended enough to be probed by space mission such as JWST. Planets d to h present condensed ice phases, with maximum water mass fractions below 20\%. 
}   
   {The derived amounts of water for TRAPPIST-1 planets show a general increase with semi-major axis, with the exception of planet d. This deviation from the trend could be due to formation mechanisms, such as migration and an enrichment of water in the region where planet d formed, or an extended CO$_{2}$-rich atmosphere.
}


   \maketitle

%

\section{Introduction}

Ongoing space missions such as CHEOPS \citep{2017chsw.confE...1B} and TESS \citep{2015JATIS...1a4003R}, and their follow-up with ground-based radial velocity telescopes, are confirming the existence of low-mass exoplanets with a wide range of densities. These densities range from the values typically inferred for the Earth or Mercury to those measured in Uranus and Neptune. The exoplanets in the former class are mainly composed of a Fe-rich core and a silicate mantle, while the latter class has a layer that is rich in volatiles. Water is the most abundant and least dense volatile after H and He \citep{2014RSPTA.37230084F}, which makes it a likely species to constitute the volatile reservoir in these planets. Several studies have investigated the interior structure and composition of water-rich planets \citep{2007Icar..191..337S,2007ApJ...669.1279S,2015A&A...577A..83D,2019PNAS..116.9723Z}, but focused mainly on its condensed phases. Nonetheless, many sub-Neptunes are close to their host star and receive enough irradiation to trigger a runaway greenhouse state in which water is present as steam. In some cases, the high pressure and temperature conditions can render the hydrosphere in supercritical and plasma, or even superionic phases \citep{2019A&A...621A.128M,2016PhRvE..93b2140F}. Therefore, it is crucial to include the modeling of all possible phases of water to provide an accurate description of its presence on the planetary surface. Moreover, the surface conditions are determined by the greenhouse effect caused by atmospheric gases, making the modelling of radiative-convective equilibrium in atmospheres a key parameter to determine in which phase water could be present on the surface. Most of interior structure models represent the planetary atmosphere as a gas layer with a simplified isothermal temperature profile \citep{2018ApJ...865...20D,2017A&A...597A..37D}, which is very different from the temperature profile in the convective deep layers of thick atmospheres \citep{2012JGRE..117.1001M}.

Multiplanetary systems are unique environments that present both planets that can hold condensed phases as well as highly-irradiated planets with steam atmospheres. In this study, we develop a planet interior model suitable for the different conditions at which water can be found in low-mass planets. Our implementation includes a supercritical water layer, introduced in \cite{2020ApJ...896L..22M}, coupled with a 1D radiative-convective atmosphere model \citep{2012JGRE..117.1001M, 2017JGRE..122.1539M,2019Icar..317..583P} to calculate the total radius of the highly-irradiated planets with water self-consistently. Furthermore, for temperate planets, we have updated the interior model presented in \cite{2016ApJ...831L..16B, 2017ApJ...850...93B} to include ice phases Ih, II, III, V and VI. We introduce these models in a MCMC Bayesian analysis scheme adapted from \cite{2015A&A...577A..83D}. This allows us to derive the water mass fraction (WMF) and core mass fraction (CMF) with their associated confidence intervals that reproduce the observed radius, mass and stellar composition measurements.

We use this model to explore the possible water content of the TRAPPIST-1 system, an ultra-cool M dwarf that hosts seven low-mass planets in close-in orbits. Three of these planets are located in the habitable zone \citep{2018A&A...613A..68G}, meaning that they can hold liquid water or ice Ih on their surfaces. Although all planets in TRAPPIST-1 system have masses and radii that are characteristic of rocky planets, their differences in density indicate that each planet has a different volatile content. This makes this planetary system ideal for testing planet interior, atmospheric structure and formation scenarios. 

In Sect.~\ref{sec:models} we describe the complete interior structure model, including the new updates for the supercritical and ice phases, the coupling between the interior and the atmosphere for steam and supercritical planets, and the MCMC Bayesian algorithm. The parameters for the TRAPPIST-1 planets used in this study are summarized in Sect.~\ref{sec:data}, including mass, radius and Fe/Si molar ratio. The results of our analysis of the hydrospheres of TRAPPIST-1 planets are described in Sect.~\ref{sec:results}. We compare our results with previous works and discuss the implications of our water estimates for planet formation in Sect.~\ref{sec:discussion}. We finally expose our conclusions in Sect.~\ref{sec:conclusion}.

\section{Planetary structure model}
\label{sec:models}

For consistency, we recall the main principles of the interior structure model. The basis of our model  is explained in \cite{2016ApJ...831L..16B, 2017ApJ...850...93B}. The 1D interior structure model takes as input the mass and the composition of the planet, which is parameterized by the CMF and WMF. The structure of the planet is stratified in 3 layers: a core, a mantle and a hydrosphere. The pressure, temperature, gravity acceleration and density are computed at each point of the one-dimensional spatial grid along the radius of the planet. The pressure, $P(r)$, is obtained by integrating the hydrostatic equilibrium (Eq.~\ref{eqn:dpdr}); the gravitational acceleration, $g(r)$, by solving Gauss's theorem (Eq.~\ref{eqn:dgdr}); the temperature, $T(r)$, with the adiabatic gradient (Eq.~\ref{eqn:dtdr}); and the density, $\rho (r)$, with the Equation of State (EOS).  $m$ is the mass at radius $r$, $G$ is the gravitational constant, and $\gamma$ and $\phi$ are the Gruneisen and the seismic parameters, respectively. Their formal macroscopic definitions are shown in equation~\ref{eqn:gruneisen}, where $E$ is the internal energy and $V$ is the volume. The Gruneisen parameter is a thermodynamic parameter that describes the dependence of the vibrational properties of a crystal with the size of its lattice. It relates the temperature in a crystalline structure to the density, which is calculated by the EOS. The seismic parameter defines how seismic waves propagate inside a material. It is related to the slope of the EOS at constant temperature \citep{2017ApJ...850...93B,2007Icar..191..337S}.

\begin{equation}
\label{eqn:dpdr}
\dfrac{dP}{dr} = - \rho g
\end{equation}

\begin{equation}
\label{eqn:dgdr}
\dfrac{dg}{dr} = 4 \pi G \rho - \dfrac{2 G m}{r^{3}}
\end{equation}

\begin{equation}
\label{eqn:dtdr}
\dfrac{dT}{dr} = - g \dfrac{\gamma T}{\phi}
\end{equation}

\begin{equation}
\label{eqn:gruneisen}
\begin{cases}
\phi = \dfrac{dP}{d \rho}  \\
\gamma = V \  \left(  \dfrac{dP}{dE} \right)_{V} 
\end{cases}
\end{equation}

The boundary conditions are the temperature and pressure at the surface, and the gravitational acceleration at the center of the planet. The value of the latter is zero. The total mass of the planet is calculated with Eq.~\ref{eqn:dmdr}, which is derived from the conservation of mass \citep{2017ApJ...850...93B,2007Icar..191..337S}. Once the total input mass of the planet is reached and the boundary conditions are fulfilled, the model has converged.

\begin{equation}
\label{eqn:dmdr}
\dfrac{dm}{dr} = 4 \pi r^{2} \rho
\end{equation}

Depending on the surface conditions, the hydrosphere can be present in supercritical, liquid or ice states. For each of these phases of water, we use a different EOS and Gruneisen parameter to compute their P-T profiles and density accurately. In Sect.~\ref{subsec:supercritical} we describe the updates to the supercritical water layer with respect to the model depicted in \cite{2020ApJ...896L..22M}, while in Sect.~\ref{subsec:ices} we present the implementation of the hydrosphere in ice phases. Finally, the coupling between the atmosphere and the interior model with planets whose hydrosphere is in steam or supercritical phases is explained in Sect.~\ref{subsec:coupling}, followed by the description of the MCMC algorithm in Sect.~\ref{subsec:mcmc}.

\subsection{Supercritical water}
\label{subsec:supercritical}

If the planet is close enough to its host star, the upper layer of the hydrosphere corresponds to a hot steam atmosphere, whose temperature at the base is determined by the radiative-convective balance calculated by the atmosphere model \citep{2012JGRE..117.1001M, 2017JGRE..122.1539M}. When the pressure and temperature at the surface, which is defined as the base of the hydrosphere layer, are above the critical point of water, we include a supercritical water layer extending from the base of the hydrosphere to a height corresponding to the phase change to steam \citep{2020ApJ...896L..22M}. We updated the EOS for this layer to the EOS introduced by \cite{2019A&A...621A.128M}, which is a fit to the experimental data provided by the International Association for the Properties of Water and Steam (IAPWS) \citep{doi:10.1063/1.1461829} for the supercritical regime, and quantum molecular dynamics (QMD) simulations data for plasma and superionic water \citep{2009PhRvB..79e4107F}.
The IAPWS experimental data span a temperature range of 251.2 to 1273 K and of 611.7 to 10$^{9}$ Pa in pressure, while their EOS can be extrapolated up to 5000 K in temperature and 10$^{11}$ Pa in pressure  \citep{doi:10.1063/1.1461829}. The validity range of the EOS presented in \cite{2019A&A...621A.128M} includes that of the IAPWS plus the region in which the QMD simulations are applicable, which corresponds to a temperature from 1000 K to 10$^{5}$ K and densities in the 1--$10^{2}$ \ g/cm$^{3}$ range. These densities are reached at high pressures, i.e., in the 10$^{9}$--10$^{12}$ Pa range. Following Eq.~\ref{eqn:dtdr}, the adiabatic gradient of the temperature is specified by the Gruneisen and the seismic parameters. These are dependent on the derivatives of the pressure with respect to the density and the internal energy (Eq.~\ref{eqn:gruneisen}). We make use of the specific internal energy and density provided by \cite{2019A&A...621A.128M} to calculate them. 

\subsection{Ice phases}
\label{subsec:ices}

\begin{table*}
\caption{EOS and reference thermal parameters for ices Ih, II, III, V and VI. This includes the reference values for the density $\rho_{0}$, the temperature $T_{0}$, the bulk modulus $K_{T_{0}}$ and its derivative $K'_{T_{0}}$, the heat capacity $C_{p}(T_{0})$, and the thermal expansion coefficient  $\alpha_{0}$. }
\label{tab:ice_data}
\centering
\begin{tabular}{cccccccc}
\hline \hline
Phase & $\rho_{0}$ [kg m$^{-3}$] & $T_{0}$ [K] & $K_{T_{0}}$ [GPa] & $K'_{T_{0}}$ & $C_{p}(T_{0})$ [J kg$^{-1}$ K$^{-1}$] & $\alpha_{0}$ [$10^{-6}$ K$^{-1}$] & References \\ \hline 
Ih & 921.0 & 248.15 & 9.50 & 5.3 & 1913.00 & 147 & 1, 8\\
II & 1169.8 & 237.65 & 14.39 & 6.0 & 2200.00 & 350 & 1, 2, 7\\
III & 1139.0 & 237.65 & 8.50 & 5.7 & 2485.55 & 405 & 3, 4, 5, 7\\
V & 1235.0 & 237.65 & 13.30 & 5.2 & 2496.63 & 233 & 1, 4, 5, 7 \\
VI & 1270.0 & 300.00 & 14.05 & 4.0 & 2590.00 & 146 & 4, 6, 7\\ \hline
\end{tabular}
\tablebib{(1)~\citet{1990JChPh..92.1909G};
(2) \citet{1995JChPh.103.9744B}; (3) \citet{1997JChPh.10710684T}; (4) \citet{2004JPCS...65.1277T};
(5) \citet{1986JChPh..84.5862S}; (6) \citet{2014JChPh.141j4505B}; (7) \citet{2010JChPh.133n4502C};
(8) \citet{2006JPCRD..35.1021F}}
\end{table*}

We extended the hydrosphere in \cite{2016ApJ...831L..16B, 2017ApJ...850...93B} with liquid and high pressure ice VII by adding 5 more condensed phases: ice Ih, II, III, V and VI. EOS for ice Ih has been developed by \cite{2006JPCRD..35.1021F} with minimization of the Gibbs potential function from the fit of experimental data. It covers all the pressure and temperature range in which water forms ice Ih.

\cite{1993JChPh..99.5369F} proposed a formalism to derive the EOS of ices II, III and V. These EOS have the form of $V=V(P,T)$, which can be found by integrating the following differential equation \citep{2004JPCS...65.1277T}:

\begin{equation}
\label{eqn:diff_ice}
\dfrac{dV}{V} = \alpha dT - \beta dP
\end{equation}

\noindent where $\alpha$ is the thermal expansion coefficient and $\beta$ the isothermal compressibility coefficient.
If the relationship between the specific volume, $V$,  and the pressure, $P$, at a constant temperature $T=T_{0}$ is determined, Eq.~\ref{eqn:diff_ice} can be integrated as:
\begin{equation}
\label{eqn:VPT}
V(P,T) = V(P,T_{0}) \ exp\left(  \int_{T_{0}}^{T}  \alpha (P,T')   dT' \right) 
\end{equation}

\cite{1993JChPh..99.5369F} proposed the following expression for the thermal expansion coefficient $\alpha$:

\begin{equation}
\label{eqn:alpha}
\alpha (P,T) = \alpha (P_{0},T)  \dfrac{\left( 1 +\dfrac{K'_{T_{0}}}{K_{T_{0}}} P \right)^{-\eta} }{\left( 1 +\dfrac{K'_{T_{0}}}{K_{T_{0}}} P_{0} \right)^{-\eta} }   = -\dfrac{1}{\rho}  \dfrac{d \rho(T)}{dT} \dfrac{\left( 1 +\dfrac{K'_{T_{0}}}{K_{T_{0}}} P \right)^{-\eta} }{\left( 1 +\dfrac{K'_{T_{0}}}{K_{T_{0}}} P_{0} \right)^{-\eta} }
\end{equation}

\noindent where $\eta$ is an adjustable parameter estimated from the fitting of experimental data. Its value is 1.0 for ice II and ice III \citep{2002JPCS...63..843L} and 7.86 for ice V \citep{1986JChPh..84.5862S}. $\rho$ is the density, $\alpha (P_{0},T)$ is the coefficient of thermal expansion at a reference pressure $P_{0}$, $K_{T_{0}}$ is the isothermal bulk modulus at the reference temperature $T_{0}$, and $K'_{T_{0}}$ is the first derivative of the isothermal bulk modulus at the reference temperature. Hence, by substituting Eq.~\ref{eqn:alpha} in Eq.~\ref{eqn:VPT} and integrating, we obtain the following EOS for high-pressure ice: 

\begin{equation}
\label{eqn:eos_fei}
V(P,T) = V(P,T_{0}) \ exp \left[  ln \left( \dfrac{ \rho(T_{0}) }{ \rho(T) }    \right)       \dfrac{\left( 1 +\dfrac{K'_{T_{0}}}{K_{T_{0}}} P \right)^{-\eta} }{\left( 1 +\dfrac{K'_{T_{0}}}{K_{T_{0}}} P_{0} \right)^{-\eta} }    \right]
\end{equation}

\noindent The final expression (Eq.~\ref{eqn:eos_fei}) requires the knowledge of the variation of the specific volume, $V(P,T_{0})$, with pressure at the reference temperature $T_{0}$. Moreover, the variation of the density with temperature, $\rho (T)$, and the bulk modulus with its derivative at the reference temperature, $K_{T_{0}}$ and $K'_{T_{0}}$, must also be provided. In Table~\ref{tab:ice_data} we specify the data and references to obtain these parameters for each ice phase. 

In the case of ice VI, we adopt the second-order Birch-Murnaghan (BM2) formulation, which is:

\begin{equation}
\label{eqn:bm2}
P = \dfrac{3}{2} K_{T_{0}} \left[  \left(  \dfrac{\rho}{\rho_{0}}  \right)^{\dfrac{7}{3}} -  \left(  \dfrac{\rho}{\rho_{0}}  \right)^{\dfrac{5}{3}}    \right],
\end{equation}

\noindent where $\rho_{0}$ is the reference density for ice VI. We also introduce a thermal correction to the density since the pressure also depends on the temperature:

\begin{equation}
\label{eqn:thermal_corr}
\rho(T) = \rho_{0} \exp \left(      \alpha_{0}   \left(  T - T_{0}    \right) \right)
\end{equation}

\noindent where $\alpha_{0}$ is the reference coefficient of thermal expansion. Interfaces between liquid and ice layers are established by phase transition functions from \cite{2010SoSyR..44..202D}.

\subsection{Interior-atmosphere coupling}
\label{subsec:coupling}

\begin{figure} 
\centering
\includegraphics[width=\hsize]{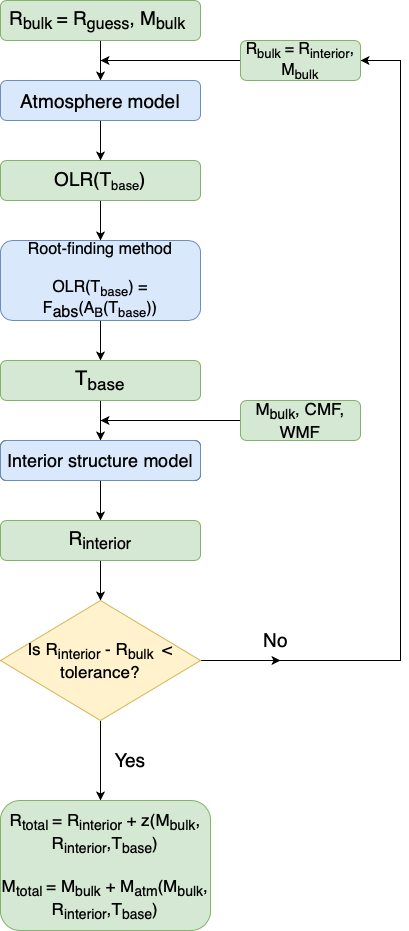}
\caption{Structural diagram of the coupling between the interior structure model and the atmosphere model. $T_{base}$ is the temperature at the bottom of the steam atmosphere in radiative-convective equilibrium. $z$ and $M_{atm}$ denote the atmospheric thickness and mass, respectively. $R_{bulk}$ and $M_{bulk}$ correspond to the planet bulk radius and mass, respectively. $R_{guess}$ refers to the initial guess of the bulk radius, while $R_{interior}$ is the output bulk radius of the interior structure model in each iteration.}
         \label{diagram}
   \end{figure}

We use a one-dimensional atmosphere model designed to compute radiative transfer and pressure-temperature ($P$, $T$) profiles for water and CO$_{2}$ atmospheres \citep{2012JGRE..117.1001M, 2017JGRE..122.1539M}. The formation of water clouds is considered in the computation of the albedo. The atmosphere is in radiative equilibrium, and presents a composition of 99\% water and 1\% CO$_{2}$. The density of steam is obtained using a non-ideal EOS \citep{osti_5614915}. 

If the surface pressure is below 300 bar, the atmosphere and the interior are coupled at the atmosphere-mantle boundary and water does not reach the supercritical regime. However, if the surface pressure is greater than 300 bar, the atmosphere and the interior are coupled at this pressure level and a layer of water in supercritical phase forms between the atmosphere and the mantle. The pressure level at 300 bar is close enough to the critical point of water at 220 bar to avoid the atmosphere model take over pressures and temperatures where the temperature profile is adiabatic.

The top-of-atmosphere pressure is set to 20 mbar, which corresponds to the observable transiting radius \citep{2020ApJ...896L..22M,2018A&A...613A..68G}. We denote the radius and mass from the center of the planet to this pressure level the total radius and mass, $R_{total}$ and $M_{total}$, respectively. We also define the radius and the mass that comprise the core, mantle and supercritical layers as the bulk radius and mass, $R_{bulk}$ and $M_{bulk}$, respectively. The atmosphere model provides the Outgoing Longwave Radiation (OLR), albedo, thickness and mass of the atmosphere as a function of the bulk mass and radius, and the surface temperature. If the atmosphere of the planet is in radiative equilibrium, the OLR is equal to the radiation the planet absorbs from its host star,  $F_{abs}$. The OLR depends on the effective temperature since OLR $= \sigma T_{\mathrm{eff}}^{4}$, where $\sigma$ corresponds to the Stefan-Boltzmann constant. To calculate the absorbed radiation $F_{abs}$, we first compute the equilibrium temperature, which is

\begin{equation} 
\label{eq:teq}
T_{\mathrm{eq}} = (1-A_{B})^{0.25} \left(  \frac{R_{\star}}{2 a_{d}}\right)^{0.5} T_{\star},
\end{equation} 

\noindent where $A_{B}$ is the planetary albedo, $R_{\star}$ and $T_{\star}$ are the radius and effective temperature of the host star, respectively. $a_{d}$ is the semi-major axis of the planet. The absorbed radiation is then calculated as

\begin{equation} 
\label{eq:rad}
F_{abs} = \sigma \ T_{\mathrm{eq}}^{4}.
\end{equation}

Figure~\ref{diagram} shows the algorithm we implemented to couple the planetary interior and the atmosphere. The interior structure model calculates the radius from the center of the planet to the base of the steam atmosphere. For a fixed set of bulk mass and radius, the OLR depends on the surface temperature. Consequently, the surface temperature at which the OLR equals the absorbed radiation corresponds to the surface temperature that yields radiative equilibrium in the atmosphere. This is estimated with a root-finding method. Since the bulk radius is an output of the interior model ($R_{interior}$) and an input of the atmosphere model, we first need to calculate the surface temperature for a certain mass and composition with an initial guess bulk radius. Then this surface temperature is the input for the interior model, which provides the bulk radius. With this bulk radius, we can generate a new value of the surface temperature. This scheme is repeated until the bulk radius converges to a constant value, to which we add the thickness of the atmosphere, $z$, to get the total radius of the planet $R_{total}$. The total mass $M_{total}$ is obtained as the sum of the bulk mass $M_{bulk}$ plus the atmospheric mass $M_{atm}$. The tolerance used to determine if the bulk radius has achieved convergence is 2\% of the bulk radius in the previous iteration. This is approximately 0.02 $R_{\oplus}$ for an Earth-sized planet.

\subsection{MCMC Bayesian analysis}
\label{subsec:mcmc}

We adapted the MCMC Bayesian analysis algorithm described in \cite{2015A&A...577A..83D} to our coupled interior and atmosphere model. The input model parameters are the bulk planetary mass $M_{bulk}$, the CMF and the WMF. Therefore, \textbf{m} = $\left\lbrace M_{bulk}, CMF, WMF \right\rbrace$, following the notation in \cite{2015A&A...577A..83D}. Depending on the planetary system and their available data, we can have observational measurements of the total planetary mass and radius and the stellar composition, or only the total planetary mass and radius. The available data in the former case is denoted as \textbf{d} = $\left\lbrace M_{obs}, R_{obs}, Fe/Si_{obs} \right\rbrace$, while the data in the latter case is represented as \textbf{d} = $\left\lbrace M_{obs}, R_{obs} \right\rbrace$. The uncertainties on the measurements are $\sigma(M_{obs}),\sigma(R_{obs})$, and $\sigma(Fe/Si_{obs})$.

The CMF and WMF prior distributions are uniform distributions between 0 and a maximum limit. This maximum limit is 75\% for the CMF, which is derived from the maximum estimated Fe/Si ratio of the proto-Sun \citep{2009LanB...4B..712L}. With this limit on the CMF, we are assuming that the exoplanets have not been exposed to events during or after their formation that could have stripped away all of their mantle, such as mantle evaporation or giant impacts. In addition, the maximum WMF is set to 80\%, which is the average water proportion found in comets in the solar system \citep{2019AJ....158..128M}. The prior distribution for the mass is a Gaussian distribution whose mean and standard deviations correspond to the central value and uncertainties of the observations.

The MCMC scheme first starts by randomly drawing a value for each of the model parameters from its respective prior distributions. This set of values is designated as \textbf{m}$_{1} = \left\lbrace M_{bulk,1}, CMF_{1}, WMF_{1} \right\rbrace $.  The index $i=1$ corresponds to the first proposed set of input values within the first chain, $n=1$. The model calculates the total mass and radius and the theoretical Fe/Si mole ratio, which are the set of output parameters g(\textbf{m}$_{1}) = \left\lbrace R_{1},M_{1}, Fe/Si_{1} \right\rbrace $. The likelihood of a set of model parameters is then calculated via the following relationship \citep{2015A&A...577A..83D}:

\begin{multline} \label{eq:likelihood_old}
L(\textbf{m}_{i} \mid \textbf{d}) = 
C \ exp \ \Biggl( -\frac{1}{2} \Biggl[
\left( \frac{(R_{i}-R_{obs})}{\sigma(R_{obs})}\right)^{2} 
+ \left( \frac{(M_{i}-M_{obs})}{\sigma(M_{obs})} \right)^{2}\\
+ \left( \frac{(Fe/Si_{i}-Fe/Si_{obs})}{\sigma(Fe/Si_{obs})} \right)^{2}
 \Biggr]  \Biggr) 
\end{multline}

\noindent where the normalization constant of the likelihood function $C$ is defined as:

\begin{equation} \label{eq:norm}
C = \dfrac{1}{(2 \pi)^{3/2} \left[ \sigma^{2}(M_{obs}) \cdot \sigma^{2}(R_{obs})   \cdot \sigma^{2}(Fe/Si_{obs})  \right]^{1/2} }.
\end{equation}

When the Fe/Si mole ratio is not available as data, the square residual term of the Fe/Si mole ratio is removed from Eq.~\ref{eq:likelihood_old}, as well as its squared uncertainty in Eq.~\ref{eq:norm}. 

Subsequently we draw a new set of input parameters, \textbf{m}$_{2} = \left\lbrace M_{bulk,2}, CMF_{2}, WMF_{2} \right\rbrace $ from the prior distributions within the same chain, $n$.  We assure that the absolute difference between the values for $i=1$ and $i=2$ is lower than a fixed step, which is the maximum size of the perturbation. This guarantees that the new state \textbf{m}$_{2}$ is uniformly bounded and centered around the old state, \textbf{m}$_{1}$. The maximum perturbation size is selected so that the acceptance rate of the MCMC, which is defined as the ratio between the number of models that are accepted and the number of proposed models, is above 20\%. After \textbf{m}$_{2} $ is chosen, the forward model calculates its corresponding output parameters and obtains their likelihood $L(\textbf{m}_{2} \mid \textbf{d})$, as shown in Eq.~\ref{eq:likelihood_old}. The acceptance probability is estimated with the log-likelihoods $l(\textbf{m} \mid \textbf{d}) = log(L(\textbf{m} \mid \textbf{d}))$ as:

\begin{equation}
P_{accept} = min \left\lbrace 1, e^{(l(\textbf{m}_{2} \mid \textbf{d}) - l(\textbf{m}_{1} \mid \textbf{d}))}  \right\rbrace 
\end{equation}

If $P_{accept}$ is greater than a number drawn from a uniform distribution between 0 and 1, \textbf{m}$_{2}$ is accepted and the chain moves to the state characterised by \textbf{m}$_{2}$, starting the next chain $n+1$. Otherwise, the chain remains in the state of \textbf{m}$_{1}$ and a different set of model parameters is proposed as \textbf{m}$_{3}$. To make sure that the posterior distributions converge and that all parameter space is explored, we run $10^4$ chains. In other words, with acceptance rates between 0.2 and 0.6, the MCMC proposes between 1.6 and 5 $\times$ $10^4$ sets of model inputs.

\section{System parameters of TRAPPIST-1}
\label{sec:data}

\cite{2020arXiv201001074A} have performed an analysis of TTVs that includes all transit data from \textit{Spitzer} since the discovery of the system. We adopt these data for the mass, radius and semi-major axis in our interior structure analysis (Table~\ref{mr-data}).

TRAPPIST-1 does not have available data regarding its chemical composition. However, the Fe/Si abundance ratio can be estimated assuming that TRAPPIST-1 presents a similar chemical composition to that of other stars of the same metallicity, age and stellar population. As proposed by \cite{2018NatAs...2..297U}, we select a sample of stars from the Hypatia Catalog \citep{2014AJ....148...54H,2016ApJS..226....4H,2017ApJ...848...34H}. We choose the set of stars by constraining the C/O mole ratio to be less than 0.8, and the stellar metallicity between -0.04 and 0.12, as this is the metallicity range calculated for TRAPPIST-1 by \cite{2017Natur.542..456G}. We discard thick disk stars since TRAPPIST-1 is likely a thin disk star. Our best-fit Gaussian to the distribution of the Fe/Si mole ratio shows a mean of 0.76 and a standard deviation of 0.12. Since this Fe/Si value is an estimate based on the chemical composition of a sample of stars that belong to the same stellar population of TRAPPIST-1, we present two scenarios for each planet. In scenario 1, the only available data are the planetary mass and radius, while scenario 2 includes the estimated stellar Fe/Si mole ratio to constrain the bulk composition.

For temperate planets that cannot have a steam atmosphere, we set the surface temperature in our interior model to their equilibrium temperatures assuming an albedo zero (Table~\ref{mr-data}). Although surface temperatures for thin atmospheres are lower than that obtained with this assumption, the dependence of the bulk radius on surface temperature for planets with condensed water is low. For example, if we assume a pure water planet of 1 $M_{\oplus}$ with a surface pressure of 1 bar, the increase in radius due to a change of surface temperature from 100 K to 360 K is 0.002 $R_{\oplus}$, which is less than 0.2\% of the total radius, i.e., 10 times less than our convergence criterion. Additionally, the atmospheres of TRAPPIST-1 planets in the habitable zone and farther are significantly thinner than those of the highly-irradiated planets. \cite{2018ApJ...867...76L} estimate thicknesses of approximately 80 km for temperate planets in TRAPPIST-1, which is negligible compared to their total radius. Therefore, we only calculate the atmospheric parameters (OLR, surface temperature, albedo and thickness of the atmosphere) for planets that present their hydrospheres in steam phase.

\begin{table*}[]
\centering
\begin{tabular}{ccccc}
\hline \hline
Planet & $M$ [$M_{\oplus}$] & $R$ [$R_{\oplus}$] & $a_{d}$ [$10^{-2}$ AU] & $T_{eq} [K]$\\ \hline
b & 1.374$\pm$0.069 & 1.116$^{+0.014}_{-0.012}$ &  1.154  & 398\\
c & 1.308$\pm$0.056 & 1.097$^{+0.014}_{-0.012}$ &  1.580 & 340\\
d & 0.388$\pm$0.012 & 0.788$^{+0.011}_{-0.010}$ &  2.227 & 286 \\
e & 0.692$\pm$0.022 & 0.920$^{+0.013}_{-0.012}$ &  2.925 & 250\\
f  & 1.039$\pm$0.031 & 1.045$^{+0.013}_{-0.012}$ & 3.849 & 218\\
g & 1.321$\pm$0.038  & 1.129$^{+0.015}_{-0.013}$ &  4.683 & 197\\
h & 0.326$\pm$0.020 &  0.775$\pm$0.014 &  6.189 & 172\\ \hline
\end{tabular}
\caption{Masses, radii and semi-major axis for all planets in TRAPPIST-1 \citep{2020arXiv201001074A}. Equilibrium temperatures are calculated assuming a null albedo, with the stellar effective temperature, stellar radius and semi-major axis provided by \cite{2020arXiv201001074A}. }
\label{mr-data}
\end{table*}

\section{Characterisation of hydrospheres}
\label{sec:results}

\subsection{CMF and WMF posterior distributions}

\begin{table*}[]
\centering
\begin{tabular}{cccccc}
\hline \hline
Planet & $M_{ret} \ [M_{\oplus}]$ & $R_{ret} \ [R_{\oplus}]$ & CMF & WMF & Fe/Si$_{ret}$  \\ \hline
b & 1.375$\pm$0.041 & 1.116$\pm$0.013 & 0.261$\pm$0.146 & (3.1$_{-3.1}^{+5.0}$) $10^{-5}$  & 1.00$\pm$0.56  \\
c & 1.300$\pm$0.036 & 1.103$\pm$0.015 & 0.239$\pm$0.084 & (0.0$_{-0.0}^{+4.4}$) $10^{-6}$ & 0.71$\pm$0.26  \\
d & 0.388$\pm$0.007 & 0.790$\pm$0.010 & 0.409$\pm$0.167 & 0.084$\pm$0.071 & 1.22$^{+1.30}_{-1.22}$ \\
e & 0.699$\pm$0.013 & 0.922$\pm$0.015 & 0.447$\pm$0.123 & 0.094$\pm$0.067 & 1.75$\pm$1.17  \\
f  & 1.043$\pm$0.019 & 1.047$\pm$0.015 & 0.409$\pm$0.140 & 0.105$\pm$0.073 & 1.44$\pm$1.14  \\
g & 1.327$\pm$0.024& 1.130$\pm$0.016 & 0.399$\pm$0.144 & 0.119$\pm$0.080 & 1.33$\pm$1.29  \\
h & 0.327$\pm$0.012 & 0.758$\pm$0.013 & 0.341$\pm$0.192 & 0.081$^{+0.089}_{-0.081}$ & 0.13$^{+1.80}_{-0.13}$  \\ \end{tabular}
\caption{Output parameters retrieved by the MCMC method for all TRAPPIST-1 planets: total mass ($M_{ret}$) and radius ($R_{ret}$), CMF, WMF and Fe/Si molar ratio. In this case the mass and radius are considered as input data (scenario 1).}
\label{TRAPPIST_all_final1}
\end{table*}

\begin{table*}[]
\centering
\begin{tabular}{ccccccc}
\hline \hline
Planet & $M_{ret} \ [M_{\oplus}]$ & $R_{ret} \ [R_{\oplus}]$ & CMF & WMF & Fe/Si$_{ret}$  \\ \hline
b & 1.359$\pm$0.043 & 1.124$\pm$0.016 & 0.259$\pm$0.032 & (0.0$_{-0.0}^{+3.4}$) $10^{-6}$  & 0.79$\pm$0.10  \\
c & 1.299$\pm$0.034 & 1.103$\pm$0.014 & 0.257$\pm$0.031 & (0.0$_{-0.0}^{+2.7}$) $10^{-6}$ & 0.79$\pm$0.11  \\
d & 0.387$\pm$0.007 & 0.792$\pm$0.010 & 0.241$\pm$0.032 & 0.036$\pm$0.028 & 0.76$\pm$0.12  \\
e & 0.695$\pm$0.012 & 0.926$\pm$0.012 & 0.249$\pm$0.031 & 0.024$_{-0.024}^{+0.027}$ & 0.78$\pm$0.12  \\
f & 1.041$\pm$0.019 & 1.048$\pm$0.013 & 0.240$\pm$0.031 & 0.037$\pm$0.026 & 0.76$\pm$0.12  \\
g & 1.331$\pm$0.023 & 1.131$\pm$0.015 & 0.235$\pm$0.031 & 0.047$\pm$0.028 & 0.75$\pm$0.12  \\
h & 0.326$\pm$0.011 & 0.758$\pm$0.013 & 0.232$\pm$0.032 & 0.055$\pm$0.037 & 0.75$\pm$0.12  \\ \hline
\end{tabular}
\caption{Output parameters retrieved by the MCMC method for all TRAPPIST-1 planets: total mass ($M_{ret}$) and radius ($R_{ret}$), CMF, WMF and Fe/Si molar ratio. In this case the Fe/Si mole ratio estimated by following \cite{2018NatAs...2..297U} is also included as data (scenario 2).}
\label{TRAPPIST_all_final2}
\end{table*}

\begin{figure}[hbt!]
\centering
\begin{subfigure}[b]{\hsize}
\centering
\includegraphics[width=\textwidth]{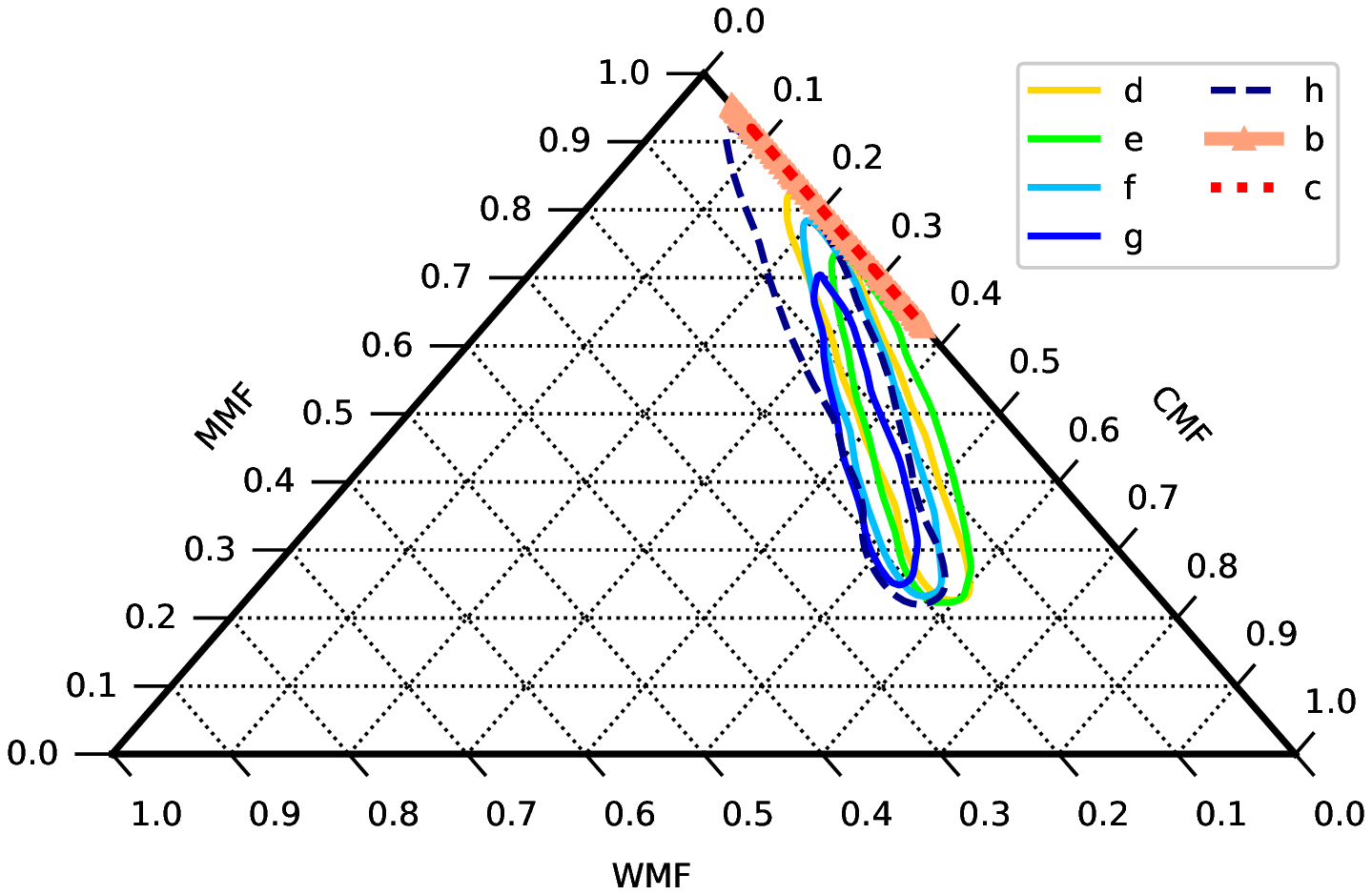}
\end{subfigure}
\hfill
\begin{subfigure}[b]{\hsize}
\centering
\includegraphics[width=\textwidth]{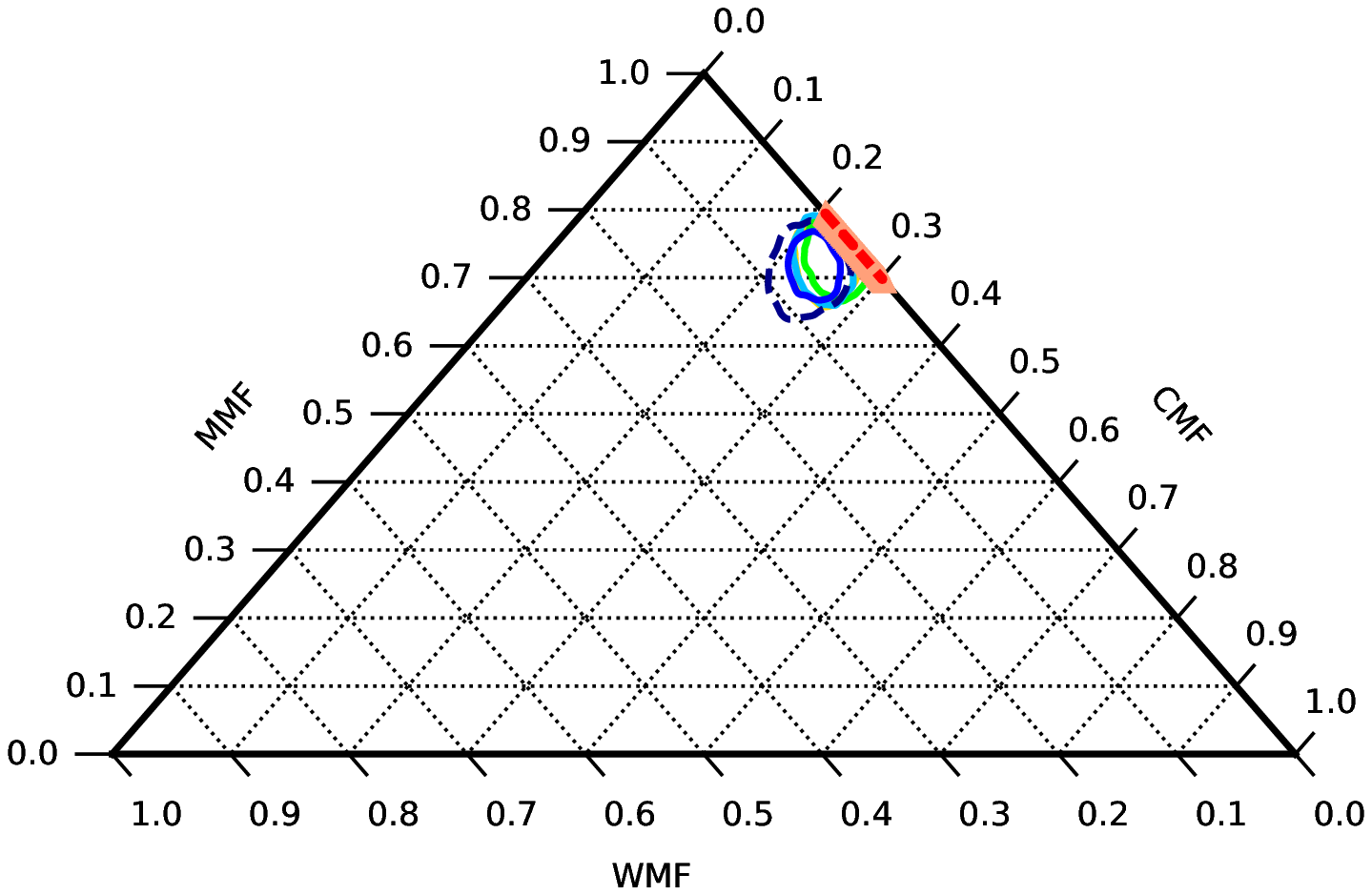}
\end{subfigure}
\caption{Top panel:1$\sigma$--confidence regions derived from the two-dimensional posterior distributions for the first scenario, where only the masses and radii are available as data. Bottom panel: 1$\sigma$ confidence regions derived from the two-dimensional posterior distributions for the second scenario, where the Fe/Si abundance ratio from \cite{2018NatAs...2..297U} is considered together with the mass and radius for each planet. The axis of the ternary diagram indicate the CMF, the WMF and the mantle mass fraction MMF = 1 - CMF - WMF.}
\label{ternary}
\end{figure}

Tables~\ref{TRAPPIST_all_final1} and ~\ref{TRAPPIST_all_final2} show the retrieved parameters, including the total planetary mass and radius, and the Fe/Si mole ratio. In both scenarios, we retrieve the mass and radius within the 1$\sigma$--confidence interval of the measurements for all planets. In scenario 1, where only the mass and radius data are considered, we retrieve Fe/Si mole ratios without any assumptions on the chemical composition of the host star. Although the uncertainties of these estimates are more than 50\% in some cases, we can estimate a common Fe/Si mole ratio for the planetary system. This common Fe/Si range is determined by the overlap of the 1$\sigma$ confidence intervals of all planets, which corresponds to Fe/Si = 0.45-0.97. This interval is compatible with the Fe/Si mole ratio of 0.76$\pm$0.12 proposed by \cite{2018NatAs...2..297U}. This overlap can also be seen in Fig.~\ref{ternary}, which presents the 1$\sigma$--confidence regions derived from the 2D marginalized posterior distributions of the CMF and WMF. The minimum value of the common CMF is determined by the lower limit of the confidence region of planet g, which is approximately 0.23, whereas the common maximum CMF value corresponds to the upper limit of planets b and c, which is 0.4. This is partially in agreement with the CMF obtained in scenario 2, when we assume the Fe/Si mole ratio proposed by \cite{2018NatAs...2..297U}, which are found between 0.2 and 0.3 (Table~\ref{TRAPPIST_all_final2}). Thus, the CMF of the TRAPPIST-1 planets could be compatible with an Earth-like CMF (CMF$_{\oplus}=$ 0.32).

In scenario 1, the retrieved WMF for all planets in the system are below 20\% within their uncertainties. This maximum WMF limit reduces to 10\% for scenario 2. This indicates that the TRAPPIST-1 system is poor in water and other volatiles, especially the inner planets b and c. Both planets are compatible with a dry composition in both scenarios, although the presence of an atmosphere cannot be ruled out given the possible CMF range estimated in scenario 1.

\subsection{Water phases}

\begin{figure}
\centering
\includegraphics[width=\hsize]{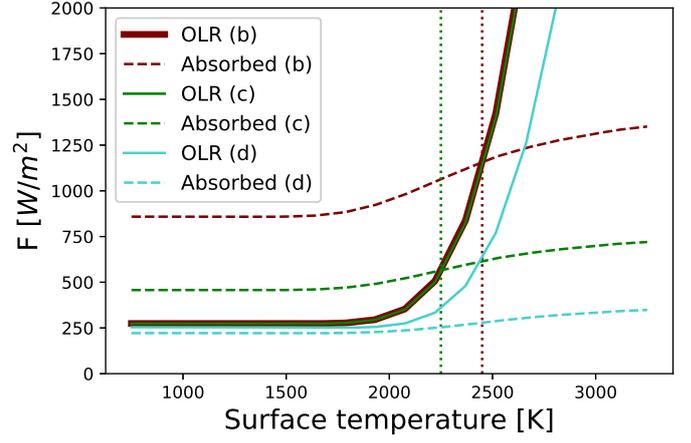}
\caption{OLR and absorbed radiation as a function of surface temperature for the steam atmospheres of TRAPPIST-1 b, c and d. Vertical dotted lines indicate the surface temperature at which the absorbed flux is equal to the OLR for planets b and c.}
\label{OLRbalance}
\end{figure}

Figure~\ref{OLRbalance} shows the OLR calculated by the atmosphere model and the absorbed radiation (Eq.~\ref{eq:teq} and~\ref{eq:rad}) for planets b, c and d. For temperatures lower than $\sim$ $T_{surf}=2000 $ K, the OLR has little dependency on the surface temperature. This is caused by the nearly constant temperature (between 250 and 300 K) of the radiating layers in the thermal IR range \citep{2013NatGe...6..661G} and it is related to the runaway greenhouse effect \citep{1969JAtS...26.1191I}. We obtain a constant OLR or an OLR limit  \citep{1992JAtS...49.2256N} of 274.3, 273.7 and 254.0 $W/m^{2}$ for planets b, c and d, respectively. These are close to the OLR limit obtained by \cite{2019ApJ...875...31K} of 279.6 $W/m^{2}$ for an Earth-like planet. The small difference is due to their different surface gravities. As explained in Sect.~\ref{subsec:coupling}, if the atmosphere can find a surface temperature at which the OLR and the absorbed radiation are equal, their atmospheres are in global radiative balance. This is the case for planets b and c, whose surface temperatures are approximately 2450 K  and 2250 K, respectively. These are above the temperatures where the blanketting effect is effective, named $T_{\varepsilon}$ in \cite{2017JGRE..122.1539M} implying that the atmospheres of planets b and c are in a post-runaway state. However, planet d is not in global radiative balance since its absorbed radiation never exceeds its OLR. This means that planet d would be cooling down, and an internal flux of approximately 33 $W/m^{2}$ would be required to supply the extra heat to balance its radiative budget. TRAPPIST-1 inner planets are likely to present an internal heat source due to tidal heating \citep{2018A&A...613A..37B,2019A&A...624A...2D,2018A&A...612A..86T}. The tidal heat flux estimated for planet d is $F_{tidal}=0.16 \ W/m^{2}$ \citep{2018A&A...613A..37B}, which is one order of magnitude lower than needed for radiative-convective balance of a steam atmosphere. Due to the blanketting effect of radiation over the surface of planet d, the OLR limit is larger than the absorbed radiation and hence the planet can cool enough to present its hydrosphere in condensed phases.

\begin{figure}
\centering
\includegraphics[width=\hsize]{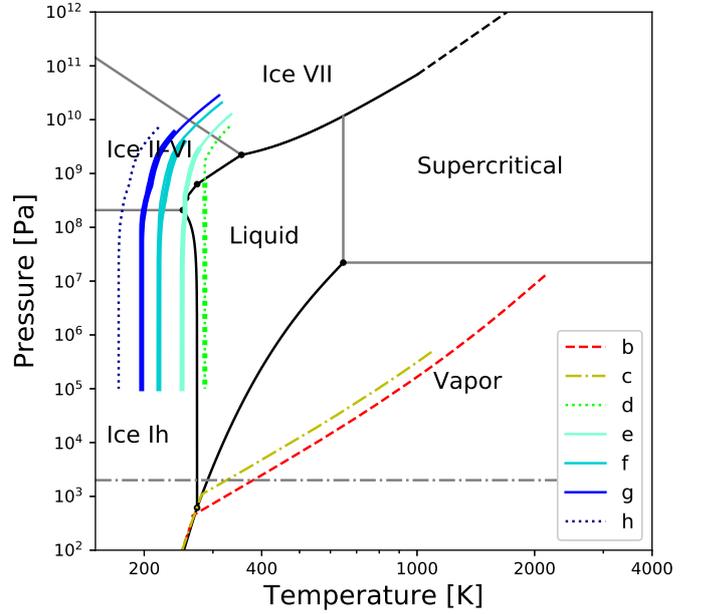}
\caption{($P$,$T$) profiles of the hydrospheres of TRAPPIST-1 planets. The dashed-dotted grey horizontal line indicates the 20 mbar pressure level (see text). Thicker lines indicate the profile for the minimum WMF estimated for each planet in scenario 1, while thinner lines mark the profile for the maximum WMF under the same scenario. The minimum WMF of planets b, c and h is zero.}
\label{waterphase}
\end{figure}

Figure~\ref{waterphase} shows the ($P$,$T$) profiles and the different phases of water we can find in the hydrospheres of the TRAPPIST-1 planets. The maximum WMF of planets b and c are $8.1 \ \times \ 10^{-5}$  and $4.4 \ \times \ 10^{-6}$, which correspond to a surface pressure of 128.9 bar and 4.85 bar, respectively. 

The thermal structure of their steam atmospheres are dominated by a lower, unsaturated troposphere where water condensation does not occur. Then the atmosphere consists of a middle, saturated troposphere where cloud formation would be possible, extending up to 10 mbar, and finally an isothermal mesosphere above. Since we consider a clear transit radius of 20 mbar \citep{2018A&A...613A..68G,2020ApJ...896L..22M} the presence of clouds above this pressure level would flatten the water features in the planetary spectrum \citep{2019A&A...628A..12T,2020A&A...643A..81K}. On the other hand, planets d and e could present water in liquid phase, which could be partially or completely covered in ice Ih. While the hydrosphere of planet h is not massive enough to attain the high pressures required for ice VII at its base, planets d, to g can reach pressures up to a 100 GPa.

\subsection{Retrieval of atmospheric parameters}

\begin{figure*}
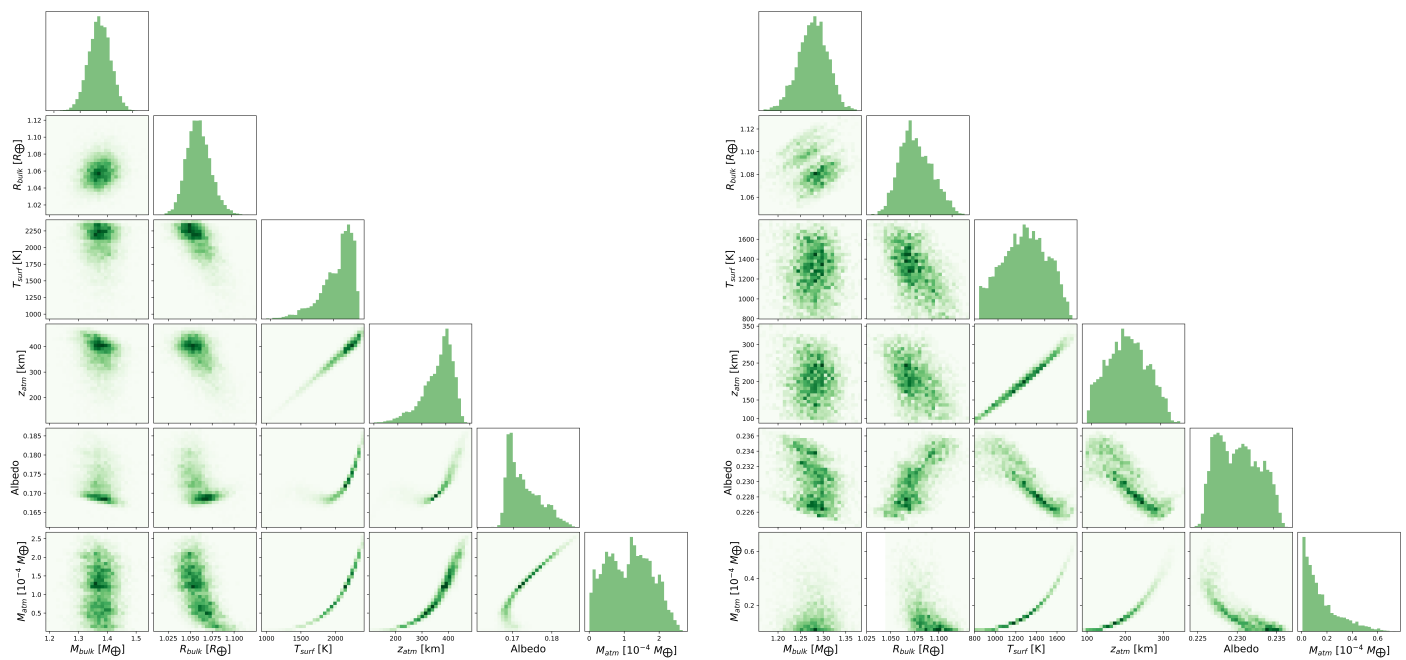
 
\centering
\begin{subfigure}[b]{0.49\textwidth}
\centering
\includegraphics[width=\textwidth]{Figures/pyramid_T_nointerp_b-2.png}
\end{subfigure}
\hfill
\begin{subfigure}[b]{0.49\textwidth}
\centering
\includegraphics[width=\textwidth]{Figures/pyramid_T_nointerp_c-2.png}
\end{subfigure}
\caption{2D and 1D marginal posterior distributions for the atmospheric parameters (surface temperature $T_{surf}$, atmospheric thickness $z_{atm}$, albedo and atmospheric mass $M_{atm}$), and bulk mass and radius, $M_{bulk}$ and $R_{bulk}$, of TRAPPIST-1 b (left panel) and c (right panel). These have been derived under scenario 1, where we do not consider Fe/Si data.}
\label{fig:atm_b_c}
\end{figure*}

Figure~\ref{fig:atm_b_c} shows the output atmospheric parameters (surface temperature, atmospheric thickness, albedo and atmospheric mass) of TRAPPIST-1 b and c for a water-dominated atmosphere in scenario 1. The total thickness of an atmosphere is related to its scale height, which is defined as $H =R T / \mu g$, where R = 8.31 J/K mol is the gas constant, $T$ is the mean atmospheric temperature, $\mu$ the mean molecular mass and $g$ the surface gravity acceleration. For planets b and c, the mean atmospheric temperatures are 940.4 and 499.4 K, and their surface gravities are 10.8 and 10.7 $m/s^{2}$, respectively. The mean molecular mass for a 99\% water and 1\% CO$_{2}$ atmosphere is 18.3 g/mol. The mean temperature increases with surface temperature, while the mean molecular mass is determined by the composition of the atmosphere. 

For the same composition and surface gravity, the scale height and therefore the thickness of the atmosphere is directly correlated to the surface temperature. As shown in Fig.~\ref{fig:atm_b_c}, the atmospheric thickness, $z_{atm}$ increases with the surface temperature $T_{surf}$. This is  known as the runaway greenhouse radius inflation effect \citep{2015AsBio..15..362G,2019A&A...628A..12T}, where a highly irradiated atmosphere is more extended than a colder one despite having similar compositions. For planet b, its atmosphere can extend up to 450 km, while planet c presents a maximum extension of 300 km. The minimum limit for the thicknesses is zero, which corresponds to the case of a dry composition. \cite{2020NatSR..1010907O} estimated that for a planet of 1-1.5 $M_{\oplus}$ the maximum atmospheric thickness due to the outgassing of an oxidised mantle is 200 km, which is compatible with the ranges we have obtained for the atmospheric thicknesses. Scenario 2 shows the same trends for the atmospheric parameters but with lower atmospheric mass and surface pressure. With their WMF posterior distributions centered in zero and low standard deviation, the surface pressure is below 1 bar and atmospheric thicknesses below 100 km in most of the accepted models, which means that in scenario 2 planets b and c are most likely dry rocky planets.

\section{Discussion}
\label{sec:discussion}

\subsection{WMF comparison with previous works}

\begin{figure}
\centering
\includegraphics[width=\hsize]{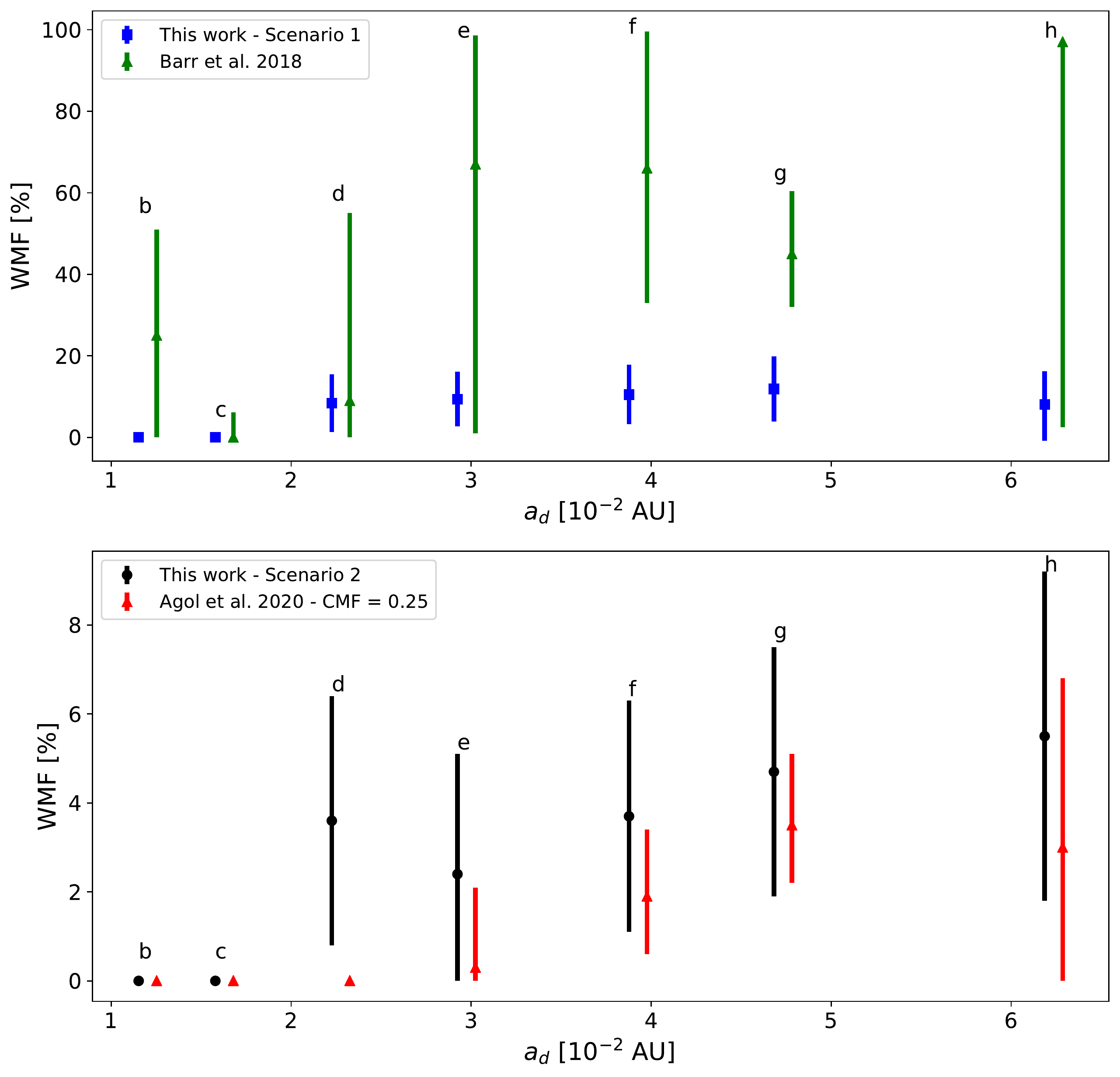}
\caption{Water mass fraction as a function of the distance to the star for the TRAPPIST-1 system. Upper panel shows our estimates for scenario 1 and those of \cite{2018A&A...613A..37B}, where only mass and radius data were taken into account. The lower panel corresponds to scenario 2, whose CMF is constrained in a narrow range between 0.2 and 0.3, while for \cite{2020arXiv201001074A} we show the WMF for a CMF of 0.25.}
\label{fig:WMF_axis}
\end{figure}

\cite{2020arXiv201001074A} use the interior and atmosphere models presented in \cite{2018ApJ...865...20D} and \cite{2020A&A...638A..41T} to obtain the WMF estimated of the TRAPPIST-1 planets with updated and more precise radii and masses data from \textit{Spitzer} TTVs \citep{2020arXiv201001074A}. We thus limited the comparison to the sole results of \cite{2020arXiv201001074A} with the same input values. By doing so, we can be certain that the variations in WMF estimates are due to our different modelling approach. Figure~\ref{fig:WMF_axis} shows that planets b and c are most likely dry in scenario 2, where the resulting CMF are between 0.2 and 0.3 for the whole system. We obtain maximum estimates of 3.4 $\times$ $10^{-6}$ and 2.7 $\times$ $10^{-6}$ for b and c, respectively. For the same density, the estimated value of the WMF depends on the CMF that is considered. Therefore we compare WMF estimates for similar CMF between this work and \cite{2020arXiv201001074A}. We show our WMF in scenario 2, since the CMF of all planets spans a narrow range between 0.2 and 0.3, which are the most similar values to one of the CMF assumed by \cite{2020arXiv201001074A}, CMF = 0.25. Our WMF for the steam planets of the TRAPPIST-1 system are in agreement with \cite{2020arXiv201001074A}, who calculated a maximum WMF of $10^{-5}$ for a constant CMF of 0.25. We are able to reduce the maximum limit of the water content of the highly irradiated planets compared to previous studies and establish the most likely WMF with our coupled atmosphere-interior model. The calculation of the total radius requires a precise determination of the atmospheric thickness. This depends strongly on the surface temperature and the surface gravity, which are obtained with radiative transfer in the atmosphere, and the calculation of the gravity profile for a bulk mass and composition in the interior self-consistently.

In the case of planet d, we estimate a WMF of 0.036 $\pm$ 0.028, while \cite{2020arXiv201001074A} obtain an upper limit of $10^{-5}$. The latter estimate considers that water is in vapor form, which is less dense than condensed phases, while our model shows that the surface conditions allow liquid or ice phases, resulting in a higher WMF. This discrepancy in the possible water phases on the surface of planet d is due to different atmospheric compositions. We consider a water-dominated atmosphere with 1\% CO$_{2}$, while \cite{2020arXiv201001074A} and \cite{2020A&A...638A..41T} assume a N$_{2}$ and H$_{2}$O mixture. This difference in composition changes radiative balance since CO$_{2}$ is a strong absorber in the IR compared to N$_{2}$, which is a neutral gas. Nonetheless, N$_{2}$ is subject to stellar wind-driven escape and it is unlikely to be stable for the inner planets of TRAPPIST-1, while CO$_{2}$ is more likely to survive thermal and ion escape processes \citep{2020SSRv..216..100T}.

Our WMF for planets e to h are in agreement within uncertainties with \cite{2020arXiv201001074A}, although their central values are significantly lower. The EOS employed to compute the density of the water layers in  \cite{2020arXiv201001074A} is also used in \cite{2018ApJ...865...20D} and \cite{2013MNRAS.434.3283V}, which agrees well with the widely-used SESAME and ANEOS EOSs \citep{2008A&A...482..315B}. These EOS are not consistent with experimental and theoretical data since they overestimate the density at pressures higher than 70 GPa \citep{2019A&A...621A.128M}. This yields an underestimation of the WMF for the same total planetary density and CMF.

For the specific case of scenario 1, with no assumptions on the stellar composition and the Fe/Si mole ratio, we compared our CMF and WMF with those obtained in \cite{2018A&A...613A..37B} (Figure~\ref{fig:WMF_axis} and Table~\ref{tab:barr}). These authors use masses and radii data given by \cite{2017arXiv170404290W}. They obtain lower masses compared to \cite{2020arXiv201001074A} while their radii are approximately similar, which would explain why \cite{2018A&A...613A..37B} tend to overestimate the water content of the TRAPPIST-1 planets. Moreover, most of the mass uncertainties in \cite{2017arXiv170404290W} are 30-50\%, while the mass uncertainties obtained by \cite{2020arXiv201001074A} are 3-5\%. This causes \cite{2018A&A...613A..37B} to calculate wider CMF and WMF 1$\sigma$ confidence intervals. In addition, there are differences between our interior modelling approach and that of \cite{2018A&A...613A..37B}. For example, according to their results, planet b can have up to 50\% of its mass as water. This high WMF value is due to the assumption that the hydrosphere is in liquid and ice I phases, and high-pressure ice polymorphs (HPPs), which are more dense than the steam atmosphere we consider. In contrast, the CMF seems to be closer to our estimates, especially for planet b, d and e, where their maximum CMF is approximately 0.40, in agreement with our CMF 1$\sigma$ intervals.

\begin{table}[]
\centering
\begin{tabular}{ccc}
\hline \hline
Planet & \multicolumn{2}{c}{CMF} \\ 
 & Barr et al. (2018) & This study (2020) \\ \hline
b & 0.00-0.43 & 0.12-0.41 \\
c & 0.00-0.98 & 0.16-0.32 \\
d & 0.00-0.39 & 0.24-0.58 \\
e & 0.00-0.40 & 0.32-0.57 \\
f & 0.00 & 0.24-0.58 \\
g & 0.00 & 0.26-0.54 \\
h & 0.00 & 0.15-0.53 \\ \hline
\end{tabular}
\caption{Comparison between our one-dimensional 1$\sigma$ confidence regions for the CMF and those of \cite{2018A&A...613A..37B}. We show only estimates for scenario 1, since \cite{2018A&A...613A..37B} did not consider any constraints on the Fe/Si ratio based on stellar composition.}
\label{tab:barr}
\end{table}

We can also discuss the possible habitability of the hydrospheres of the TRAPPIST-1 planets by comparing our WMF estimates with the layer structure as a function of planetary mass and water content obtained by \cite{2016Icar..277..215N} . According to \cite{2016Icar..277..215N}, a habitable hydrosphere must be structured in a single liquid water ocean or in several ice layers that enable the formation of a lower ocean layer. This lower ocean would be formed by the heat supplied by the mantle that melts the high pressure ice in the ice-mantle boundary \citep{2016Icar..277..215N}. For planet d, a surface liquid ocean would form for all its possible WMF if the atmosphere allows for the presence of condensed phases. For planets e, f and g, the hydrosphere could be stratified in a surface layer of ice Ih and a liquid or an ice II-VI layer. In the case we had low-pressure ices II-VI, their base could be melted by the heat provided by the mantle, and form a lower ocean layer as suggested by \cite{2016Icar..277..215N}. At WMF $\geq$ 0.10, less than 50\% of the possible configurations enable a habitable sub-surface ocean layer, and at a WMF $\geq$ 0.14, the hydrosphere is uninhabitable. In scenario 1, planets e to g reach these values within uncertainties, although their minimum values extend down to 0-0.03 in WMF, which would be the habitable regime.

\subsection{System formation and architecture}

\begin{figure}
\centering
\includegraphics[width=\hsize]{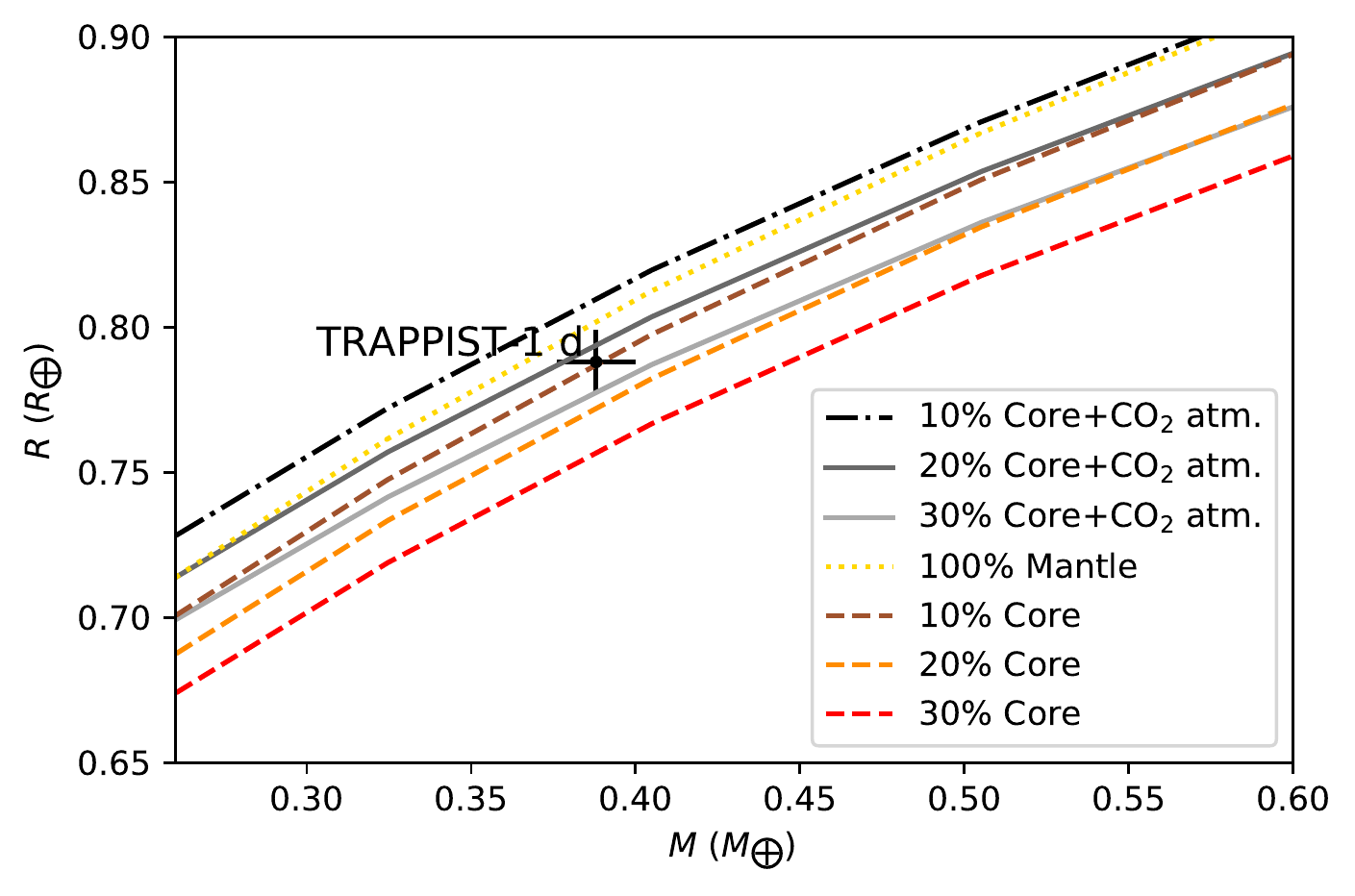}
\caption{Mass-radius relationships for planets with CO$_{2}$-dominated atmospheres assuming different CMF. The  surface pressure is 300 bar. The black dot and its error bars indicate the location and uncertainties of planet d in the mass-radius diagram. }
\label{MR_CO2}
\end{figure}

In the case of scenario 1, where no Fe/Si data is assumed, the WMF increases with the distance to the star with the exception of planet h, whose WMF is similar to that of planet d. In the case of scenario 2, where a common Fe/Si of 0.76 $\pm$ 0.12 is assumed for the whole system, the WMF increases with the distance to the star (Fig.~\ref{fig:WMF_axis}) with the exception of planet d whose WMF is similar to that of planet f, which is more water-rich than planet e. This slight deviation from the observed trend could be explained by migration, where planet d could have formed beyond the snow line and then migrated inwards \citep{2018MNRAS.479L..81R}. In addition, pebble ablation and water recycling back into the disk could have been less efficient in the case of planet d, compared to planet e \citep{2019A&A...631A...7C}. On the other hand, the gas at the distance at which planet d formed could have been more enriched in volatiles than the outer planets, accreting more water ice than planet e in a 'cold finger' \citep{1988Icar...75..146S,1998Icar..135..537C}. Pebble formation in the vicinity of the water iceline can induce important enhancements of the water ice fraction in those pebbles due to the backward diffusion of vapor through the snowline and the inward drift of ice particles. Therefore, if a planet forms from this material, it should be more water-rich than those formed further \citep{2019ApJ...875....9M}. These formation scenarios could explain the high WMF of planet d when we assume that its water layer is in condensed phases. Post-formation processes could also have shaped the trend of the WMF with axis, such as atmospheric escape due to XUV and X-ray emission from their host star. \cite{2017MNRAS.464.3728B} estimated a maximum water loss of 15 Earth Oceans (EO) for TRAPPIST-1 b and c and 1 EO for planet d. If we assume that the current WMF are the central values of the posterior distributions we derived in scenario 1, planets b, c and d would have had an initial WMF of 2.37 $\times \ 10^{-3}$, 2.50 $\times 10^{-3}$ and 0.085, respectively. Therefore, atmospheric escape would have decreased the individual WMF of each planet, but the increase of WMF with distance from the star would have been preserved.

In addition to the WMF-axis trend, we can differentiate the very water-poor, close-in planets, b and c, from the outer, water-rich planets, d to h. This has been reported as a consequence of pebble accretion in the formation of other systems, such as the Galilean moons. While Io is dry, Callisto and Ganymede are water-rich, with Europa showing an intermediate WMF of 8\% \citep{2017ApJ...845...92R}. Pebble-driven formation can produce planets with WMF $\geq 15$\% if these are formed at the water ice line \citep{2019A&A...631A...7C,2019A&A...627A.149S}. In contrast, planets formed within the ice line would present WMF less than 5\% \citep{2020A&A...638A..88L,2019A&A...631A...7C}, which is close to the mean value we calculated for planet h, 5.5\%. The maximum WMF limit in the first scenario is approximately 20\%. This maximum limit is significantly lower than the typical WMF generated by planetesimal accretion scenario, which is 50-40\% \citep{2020MNRAS.491.1998M}. Therefore, our results are consistent with the pebble-driven formation scenario.

\subsection{Alternative atmospheric compositions}
\label{sec:CO2_d}

However, the atmosphere of planet d could be dominated by other atmospheric gases different from H$_{2}$O-based mixtures, which could produce an extended atmosphere and increase the total planetary radius. Hydrogen-dominated atmospheres have been deemed unlikely as one of the possible atmospheric compositions for all planets in the TRAPPIST-1 system, both cloud-free \citep{2016Natur.537...69D,2018NatAs...2..214D} and with high-altitude clouds and hazes \citep{2018A&A...613A..68G,2020arXiv200613826D}. Similarly, CH$_{4}$-dominated atmospheres are not probable given the photometry data of the \textit{Spitzer} Space Telescope \citep{2020arXiv200613826D}. Therefore, our best candidate to explain the low density of planet d in a water-poor scenario is CO$_{2}$. We find that a CO$_2$-dominated atmosphere with 1\% water vapor in planet d would be in radiative-convective equilibrium by computing the OLR and absorbed radiation, as we did for water-dominated atmospheres. The resulting surface temperature is approximately 950 K, which is slightly higher than the surface temperature of Venus (700 K) with a higher water vapor mixing ratio. Figure~\ref{MR_CO2} introduces the mass-radius relationships for different CMF, assuming a CO$_{2}$-dominated atmosphere with a surface pressure of 300 bar. It shows that planet d is compatible with a planet with a CO$_{2}$-dominated atmosphere and CMF between 0.2 and 0.3, which is a very likely CMF range for TRAPPIST-1 planets based on our analysis. Surface pressures lower than 300 bar would yield lower atmospheric thicknesses, so it would be necessary to consider a lower CMF to explain the observed density of planet d. CO$_{2}$ in the case of planet d can be provided by volcanic outgassing \citep{2020NatSR..1010907O}, since its internal heat flux produced by tidal heating is in the range 0.04-2 $W/m^{2}$, which favours plate tectonics \citep{2018MNRAS.476.5032P}. Secondary CO$_{2}$-dominated atmospheres could have traces of O$_{2}$, N$_{2}$ and water vapor.

\section{Conclusions}
\label{sec:conclusion}

We presented an interior structure model for low-mass planets at different irradiations that is valid for a wide range of water phases, derived from the approaches of \cite{2017ApJ...850...93B} and \cite{2020ApJ...896L..22M}. For highly-irradiated planets, we couple a 1D water steam atmosphere in radiative-convective equilibrium with a high-pressure convective layer in supercritical phase. The density in this layer is computed by using an accurate EOS for high-pressure and high-temperature water phases. For temperate planets whose surface conditions allow the formation of condensed phases, we implemented a hydrosphere with liquid water and ice phases Ih, II, III, V, VI and VII. In addition, we adapted the MCMC Bayesian algorithm described in \cite{2015A&A...577A..83D} to our interior model to derive the posterior distributions of the compositional parameters, WMF and CMF, given mass, radius and stellar composition data. We then applied our interior model to the particular case of TRAPPIST-1 planets using the latest mass and radius data from \textit{Spitzer} \citep{2020arXiv201001074A}.

The hydrospheres of TRAPPIST-1 planets have been characterised by calculating their \textit{P-T} profiles and thermodynamic phases. Planets b and c are warm enough to present steam atmospheres. They could hold post-runaway greenhouse atmospheres with  thicknesses up to 450 km and surface temperatures up to 2500 K, which means that they are extended enough to be suitable targets for atmospheric characterisation by future space-based facilities such as JWST. Moreover, planets d to g present their hydrospheres in condensed phases. These hydrospheres can contain high-pressure ices that start to form at $10^{9}-10^{10}$ Pa.

We have obtained CMF and WMF probability distributions for all planets in the system. We found that the Fe/Si mole ratio of the system is in the 0.45-0.97 range without considering any assumption on the chemical composition of the stellar host. This Fe/Si range corresponds to a CMF value in the 0.23-0.40 range, making the CMF of TRAPPIST-1 planets compatible with an Earth-like value (0.32). In addition, our WMF estimates agree within uncertainties with those derived by \cite{2020arXiv201001074A}, although their most likely values are considerably lower for planets with condensed phases. In the case of planets with steam hydrospheres, their densities are compatible with dry rocky planets with no atmospheres. Nevertheless, we cannot rule out the presence of an atmosphere with the Fe/Si range we have derived without any assumption on the chemical composition of the host star. When considering a possible estimate of the Fe/Si ratio of the host star (scenario 2), we obtained lower maximum limits of the WMF for planets b and c compared to previously calculated limits by \cite{2020arXiv201001074A} for a similar CMF of 0.25. Our estimated WMF in steam and condensed phases are consistent with an increase of WMF with progressing distance from the host star. This trend, as well as the maximum WMF we calculated, favour pebble-driven accretion as a plausible formation mechanism for the TRAPPIST-1 system. However, planet d presents a slightly higher WMF than planet e. This could be due to processes that took place during planet formation, such as migration, a low-efficient ablation of pebbles and gas recycling or an enhancement of the water ice fraction in pebbles at the distance of the disc where planet d formed. An extended atmosphere dominated by greenhouse gases different from a water-dominated atmosphere, such as CO$_{2}$, could also explain the low-density of planet d compared to planet e.

Future work should include more atmospheric processes and species that determine the mass-radius relations of planets with secondary atmospheres in the Super-Earth and sub-Neptune regime. These can vary the atmospheric thickness and increase the total planetary radius with varying atmospheric masses while other compositional parameters change the bulk radius. These should be integrated in one single interior-atmosphere model, combined in a MCMC Bayesian framework such as the one we used in this study. This statistical approach has been employed with interior models for planets with H/He-dominated atmospheres \citep{2017A&A...597A..37D,2017A&A...597A..38D,2018ApJ...865...20D}, or dry planets \citep{2020MNRAS.499..932P}, but not for planets with secondary, CO$_{2}$ and steam-dominated atmospheres. The integrated model should also include a description of escape processes, such as hydrodynamic or Jeans escapes, which is particularly interesting to explore the lifetime of secondary atmospheres. Close-in, low-mass planets are likely to outgas atmospheric species, such as CO$_{2}$, and form O$_{2}$ via photodissociation of outgased H$_{2}$O, during their magma ocean stage or due to plate tectonics \citep{2020arXiv201207337C}. Thus, a mixture of these gases should be considered to study the thermal structure of planets with secondary atmospheres. Planets b and c in the TRAPPIST-1 system could present magma oceans due to their high surface temperatures (T $\geq$ 1300 K) \citep{2018A&A...613A..37B,2020arXiv201207337C}, and the maximum surface pressure we have obtained here can be used to assess the current outgassing rate in magma ocean studies (\cite{2017PEPI..269...40N}, Baumeister et al. submitted) and better constrain the WMF for the interior magma ocean models (e.g \cite{2020A&A...643A..81K}) in the future.

\begin{acknowledgements}

MD and OM acknowledge support from CNES. We acknowledge the anonymous referee whose comments helped improve and clarify this manuscript.

\end{acknowledgements}

%
%

\bibliographystyle{aa}           
\bibliography{example}      

\begin{thebibliography}{76}
\expandafter\ifx\csname natexlab\endcsname\relax\def\natexlab#1{#1}\fi

\bibitem[{{Agol} {et~al.}(2020){Agol}, {Dorn}, {Grimm}, {Turbet}, {Ducrot},
  {Delrez}, {Gillon}, {Demory}, {Burdanov}, {Barkaoui}, {Benkhaldoun},
  {Bolmont}, {Burgasser}, {Carey}, {de Wit}, {Fabrycky}, {Foreman-Mackey},
  {Haldemann}, {Hernandez}, {Ingalls}, {Jehin}, {Langford}, {Leconte},
  {Lederer}, {Luger}, {Malhotra}, {Meadows}, {Morris}, {Pozuelos}, {Queloz},
  {Raymond}, {Selsis}, {Sestovic}, {Triaud}, \& {Van
  Grootel}}]{2020arXiv201001074A}
{Agol}, E., {Dorn}, C., {Grimm}, S.~L., {et~al.} 2020, arXiv e-prints,
  arXiv:2010.01074

\bibitem[{{B{\'a}ez} \& {Clancy}(1995)}]{1995JChPh.103.9744B}
{B{\'a}ez}, L.~A. \& {Clancy}, P. 1995, \jcp, 103, 9744

\bibitem[{{Baraffe} {et~al.}(2008){Baraffe}, {Chabrier}, \&
  {Barman}}]{2008A&A...482..315B}
{Baraffe}, I., {Chabrier}, G., \& {Barman}, T. 2008, \aap, 482, 315

\bibitem[{{Barr} {et~al.}(2018){Barr}, {Dobos}, \&
  {Kiss}}]{2018A&A...613A..37B}
{Barr}, A.~C., {Dobos}, V., \& {Kiss}, L.~L. 2018, \aap, 613, A37

\bibitem[{{Benz}(2017)}]{2017chsw.confE...1B}
{Benz}, W. 2017, in CHEOPS Fifth Science Workshop, 1

\bibitem[{{Bezacier} {et~al.}(2014){Bezacier}, {Journaux}, {Perrillat},
  {Cardon}, {Hanfland }, \& {Daniel}}]{2014JChPh.141j4505B}
{Bezacier}, L., {Journaux}, B., {Perrillat}, J.-P., {et~al.} 2014, \jcp, 141,
  104505

\bibitem[{{Bolmont} {et~al.}(2017){Bolmont}, {Selsis}, {Owen}, {Ribas},
  {Raymond}, {Leconte}, \& {Gillon}}]{2017MNRAS.464.3728B}
{Bolmont}, E., {Selsis}, F., {Owen}, J.~E., {et~al.} 2017, \mnras, 464, 3728

\bibitem[{{Brugger} {et~al.}(2017){Brugger}, {Mousis}, {Deleuil}, \&
  {Deschamps}}]{2017ApJ...850...93B}
{Brugger}, B., {Mousis}, O., {Deleuil}, M., \& {Deschamps}, F. 2017, \apj, 850,
  93

\bibitem[{{Brugger} {et~al.}(2016){Brugger}, {Mousis}, {Deleuil}, \&
  {Lunine}}]{2016ApJ...831L..16B}
{Brugger}, B., {Mousis}, O., {Deleuil}, M., \& {Lunine}, J.~I. 2016, \apjl,
  831, L16

\bibitem[{{Chao} {et~al.}(2020){Chao}, {deGraffenried}, {Lach}, {Nelson},
  {Truax}, \& {Gaidos}}]{2020arXiv201207337C}
{Chao}, K.-H., {deGraffenried}, R., {Lach}, M., {et~al.} 2020, arXiv e-prints,
  arXiv:2012.07337

\bibitem[{{Choukroun} \& {Grasset}(2010)}]{2010JChPh.133n4502C}
{Choukroun}, M. \& {Grasset}, O. 2010, \jcp, 133, 144502

\bibitem[{{Coleman} {et~al.}(2019){Coleman}, {Leleu}, {Alibert}, \&
  {Benz}}]{2019A&A...631A...7C}
{Coleman}, G.~A.~L., {Leleu}, A., {Alibert}, Y., \& {Benz}, W. 2019, \aap, 631,
  A7

\bibitem[{{Cyr} {et~al.}(1998){Cyr}, {Sears}, \&
  {Lunine}}]{1998Icar..135..537C}
{Cyr}, K.~E., {Sears}, W.~D., \& {Lunine}, J.~I. 1998, \icarus, 135, 537

\bibitem[{{de Wit} {et~al.}(2016){de Wit}, {Wakeford}, {Gillon}, {Lewis},
  {Valenti}, {Demory}, {Burgasser}, {Burdanov}, {Delrez}, {Jehin}, {Lederer},
  {Queloz}, {Triaud}, \& {Van Grootel}}]{2016Natur.537...69D}
{de Wit}, J., {Wakeford}, H.~R., {Gillon}, M., {et~al.} 2016, \nat, 537, 69

\bibitem[{{de Wit} {et~al.}(2018){de Wit}, {Wakeford}, {Lewis}, {Delrez},
  {Gillon}, {Selsis}, {Leconte}, {Demory}, {Bolmont}, {Bourrier}, {Burgasser},
  {Grimm}, {Jehin}, {Lederer}, {Owen}, {Stamenkovi{\'c}}, \&
  {Triaud}}]{2018NatAs...2..214D}
{de Wit}, J., {Wakeford}, H.~R., {Lewis}, N.~K., {et~al.} 2018, Nature
  Astronomy, 2, 214

\bibitem[{{Dobos} {et~al.}(2019){Dobos}, {Barr}, \&
  {Kiss}}]{2019A&A...624A...2D}
{Dobos}, V., {Barr}, A.~C., \& {Kiss}, L.~L. 2019, \aap, 624, A2

\bibitem[{{Dorn} {et~al.}(2017{\natexlab{a}}){Dorn}, {Hinkel}, \&
  {Venturini}}]{2017A&A...597A..38D}
{Dorn}, C., {Hinkel}, N.~R., \& {Venturini}, J. 2017{\natexlab{a}}, \aap, 597,
  A38

\bibitem[{{Dorn} {et~al.}(2015){Dorn}, {Khan}, {Heng}, {Connolly}, {Alibert},
  {Benz}, \& {Tackley}}]{2015A&A...577A..83D}
{Dorn}, C., {Khan}, A., {Heng}, K., {et~al.} 2015, \aap, 577, A83

\bibitem[{{Dorn} {et~al.}(2018){Dorn}, {Mosegaard}, {Grimm}, \&
  {Alibert}}]{2018ApJ...865...20D}
{Dorn}, C., {Mosegaard}, K., {Grimm}, S.~L., \& {Alibert}, Y. 2018, \apj, 865,
  20

\bibitem[{{Dorn} {et~al.}(2017{\natexlab{b}}){Dorn}, {Venturini}, {Khan},
  {Heng}, {Alibert}, {Helled}, {Rivoldini}, \& {Benz}}]{2017A&A...597A..37D}
{Dorn}, C., {Venturini}, J., {Khan}, A., {et~al.} 2017{\natexlab{b}}, \aap,
  597, A37

\bibitem[{{Ducrot} {et~al.}(2020){Ducrot}, {Gillon}, {Delrez}, {Agol},
  {Rimmer}, {Turbet}, {G{\"u}nther}, {Demory}, {Triaud}, {Bolmont},
  {Burgasser}, {Carey}, {Ingalls}, {Jehin}, {Leconte}, {Lederer}, {Queloz},
  {Raymond}, {Selsis}, {Van Grootel}, \& {de Wit}}]{2020arXiv200613826D}
{Ducrot}, E., {Gillon}, M., {Delrez}, L., {et~al.} 2020, arXiv e-prints,
  arXiv:2006.13826

\bibitem[{{Dunaeva} {et~al.}(2010){Dunaeva}, {Antsyshkin}, \&
  {Kuskov}}]{2010SoSyR..44..202D}
{Dunaeva}, A.~N., {Antsyshkin}, D.~V., \& {Kuskov}, O.~L. 2010, Solar System
  Research, 44, 202

\bibitem[{{Fei} {et~al.}(1993){Fei}, {Mao}, \& {Hemley}}]{1993JChPh..99.5369F}
{Fei}, Y., {Mao}, H.-K., \& {Hemley}, R.~J. 1993, \jcp, 99, 5369

\bibitem[{{Feistel} \& {Wagner}(2006)}]{2006JPCRD..35.1021F}
{Feistel}, R. \& {Wagner}, W. 2006, Journal of Physical and Chemical Reference
  Data, 35, 1021

\bibitem[{{Forget} \& {Leconte}(2014)}]{2014RSPTA.37230084F}
{Forget}, F. \& {Leconte}, J. 2014, Philosophical Transactions of the Royal
  Society of London Series A, 372, 20130084

\bibitem[{{French} {et~al.}(2016){French}, {Desjarlais}, \&
  {Redmer}}]{2016PhRvE..93b2140F}
{French}, M., {Desjarlais}, M.~P., \& {Redmer}, R. 2016, \pre, 93, 022140

\bibitem[{{French} {et~al.}(2009){French}, {Mattsson}, {Nettelmann}, \&
  {Redmer}}]{2009PhRvB..79e4107F}
{French}, M., {Mattsson}, T.~R., {Nettelmann}, N., \& {Redmer}, R. 2009, \prb,
  79, 054107

\bibitem[{{Gagnon} {et~al.}(1990){Gagnon}, {Kiefte}, {Clouter}, \&
  {Whalley}}]{1990JChPh..92.1909G}
{Gagnon}, R.~E., {Kiefte}, H., {Clouter}, M.~J., \& {Whalley}, E. 1990, \jcp,
  92, 1909

\bibitem[{{Gillon} {et~al.}(2017){Gillon}, {Triaud}, {Demory}, {Jehin}, {Agol},
  {Deck}, {Lederer}, {de Wit}, {Burdanov}, {Ingalls}, {Bolmont}, {Leconte},
  {Raymond}, {Selsis}, {Turbet}, {Barkaoui}, {Burgasser}, {Burleigh}, {Carey},
  {Chaushev}, {Copperwheat}, {Delrez}, {Fernand es}, {Holdsworth}, {Kotze},
  {Van Grootel}, {Almleaky}, {Benkhaldoun}, {Magain}, \&
  {Queloz}}]{2017Natur.542..456G}
{Gillon}, M., {Triaud}, A. H.~M.~J., {Demory}, B.-O., {et~al.} 2017, \nat, 542,
  456

\bibitem[{{Goldblatt}(2015)}]{2015AsBio..15..362G}
{Goldblatt}, C. 2015, Astrobiology, 15, 362

\bibitem[{{Goldblatt} {et~al.}(2013){Goldblatt}, {Robinson}, {Zahnle}, \&
  {Crisp}}]{2013NatGe...6..661G}
{Goldblatt}, C., {Robinson}, T.~D., {Zahnle}, K.~J., \& {Crisp}, D. 2013,
  Nature Geoscience, 6, 661

\bibitem[{{Grimm} {et~al.}(2018){Grimm}, {Demory}, {Gillon}, {Dorn}, {Agol},
  {Burdanov}, {Delrez}, {Sestovic}, {Triaud}, {Turbet}, {Bolmont}, {Caldas},
  {de Wit}, {Jehin}, {Leconte}, {Raymond}, {Van Grootel}, {Burgasser}, {Carey},
  {Fabrycky}, {Heng}, {Hernandez}, {Ingalls}, {Lederer}, {Selsis}, \&
  {Queloz}}]{2018A&A...613A..68G}
{Grimm}, S.~L., {Demory}, B.-O., {Gillon}, M., {et~al.} 2018, \aap, 613, A68

\bibitem[{Haar {et~al.}(1984)Haar, Gallagher, \& Kell}]{osti_5614915}
Haar, L., Gallagher, J.~S., \& Kell, G.~S. 1984

\bibitem[{{Hinkel} {et~al.}(2017){Hinkel}, {Mamajek}, {Turnbull}, {Osby},
  {Shkolnik}, {Smith}, {Klimasewski}, {Somers}, \&
  {Desch}}]{2017ApJ...848...34H}
{Hinkel}, N.~R., {Mamajek}, E.~E., {Turnbull}, M.~C., {et~al.} 2017, \apj, 848,
  34

\bibitem[{{Hinkel} {et~al.}(2014){Hinkel}, {Timmes}, {Young}, {Pagano}, \&
  {Turnbull}}]{2014AJ....148...54H}
{Hinkel}, N.~R., {Timmes}, F.~X., {Young}, P.~A., {Pagano}, M.~D., \&
  {Turnbull}, M.~C. 2014, \aj, 148, 54

\bibitem[{{Hinkel} {et~al.}(2016){Hinkel}, {Young}, {Pagano}, {Desch}, {Anbar},
  {Adibekyan}, {Blanco-Cuaresma}, {Carlberg}, {Delgado Mena}, {Liu},
  {Nordlander}, {Sousa}, {Korn}, {Gruyters}, {Heiter}, {Jofr{\'e}}, {Santos},
  \& {Soubiran}}]{2016ApJS..226....4H}
{Hinkel}, N.~R., {Young}, P.~A., {Pagano}, M.~D., {et~al.} 2016, \apjs, 226, 4

\bibitem[{{Ingersoll}(1969)}]{1969JAtS...26.1191I}
{Ingersoll}, A.~P. 1969, Journal of Atmospheric Sciences, 26, 1191

\bibitem[{{Katyal} {et~al.}(2019){Katyal}, {Nikolaou}, {Godolt}, {Grenfell},
  {Tosi}, {Schreier}, \& {Rauer}}]{2019ApJ...875...31K}
{Katyal}, N., {Nikolaou}, A., {Godolt}, M., {et~al.} 2019, \apj, 875, 31

\bibitem[{{Katyal} {et~al.}(2020){Katyal}, {Ortenzi}, {Lee Grenfell}, {Noack},
  {Sohl}, {Godolt}, {Garc{\'\i}a Mu{\~n}oz}, {Schreier}, {Wunderlich}, \&
  {Rauer}}]{2020A&A...643A..81K}
{Katyal}, N., {Ortenzi}, G., {Lee Grenfell}, J., {et~al.} 2020, \aap, 643, A81

\bibitem[{{Leon} {et~al.}(2002){Leon}, {Rodrguez Romo}, \&
  {Tchijov}}]{2002JPCS...63..843L}
{Leon}, G.~C., {Rodrguez Romo}, S., \& {Tchijov}, V. 2002, Journal of Physics
  and Chemistry of Solids, 63, 843

\bibitem[{{Lincowski} {et~al.}(2018){Lincowski}, {Meadows}, {Crisp},
  {Robinson}, {Luger}, {Lustig-Yaeger}, \& {Arney}}]{2018ApJ...867...76L}
{Lincowski}, A.~P., {Meadows}, V.~S., {Crisp}, D., {et~al.} 2018, \apj, 867, 76

\bibitem[{{Liu} {et~al.}(2020){Liu}, {Lambrechts}, {Johansen}, {Pascucci}, \&
  {Henning}}]{2020A&A...638A..88L}
{Liu}, B., {Lambrechts}, M., {Johansen}, A., {Pascucci}, I., \& {Henning}, T.
  2020, \aap, 638, A88

\bibitem[{{Lodders} {et~al.}(2009){Lodders}, {Palme}, \&
  {Gail}}]{2009LanB...4B..712L}
{Lodders}, K., {Palme}, H., \& {Gail}, H.~P. 2009, Landolt B\&ouml;rnstein, 4B,
  712

\bibitem[{{Marcq}(2012)}]{2012JGRE..117.1001M}
{Marcq}, E. 2012, Journal of Geophysical Research (Planets), 117, E01001

\bibitem[{{Marcq} {et~al.}(2017){Marcq}, {Salvador}, {Massol}, \&
  {Davaille}}]{2017JGRE..122.1539M}
{Marcq}, E., {Salvador}, A., {Massol}, H., \& {Davaille}, A. 2017, Journal of
  Geophysical Research (Planets), 122, 1539

\bibitem[{{Mazevet} {et~al.}(2019){Mazevet}, {Licari}, {Chabrier}, \&
  {Potekhin}}]{2019A&A...621A.128M}
{Mazevet}, S., {Licari}, A., {Chabrier}, G., \& {Potekhin}, A.~Y. 2019, \aap,
  621, A128

\bibitem[{{McKay} {et~al.}(2019){McKay}, {DiSanti}, {Kelley}, {Knight},
  {Womack}, {Wierzchos}, {Harrington Pinto}, {Bonev}, {Villanueva}, {Dello
  Russo}, {Cochran}, {Biver}, {Bauer}, {Vervack}, {Gibb}, {Roth}, \&
  {Kawakita}}]{2019AJ....158..128M}
{McKay}, A.~J., {DiSanti}, M.~A., {Kelley}, M. S.~P., {et~al.} 2019, \aj, 158,
  128

\bibitem[{{Miguel} {et~al.}(2020){Miguel}, {Cridland}, {Ormel}, {Fortney}, \&
  {Ida}}]{2020MNRAS.491.1998M}
{Miguel}, Y., {Cridland}, A., {Ormel}, C.~W., {Fortney}, J.~J., \& {Ida}, S.
  2020, \mnras, 491, 1998

\bibitem[{{Mousis} {et~al.}(2020){Mousis}, {Deleuil}, {Aguichine}, {Marcq},
  {Naar}, {Aguirre}, {Brugger}, \& {Gon{\c{c}}alves}}]{2020ApJ...896L..22M}
{Mousis}, O., {Deleuil}, M., {Aguichine}, A., {et~al.} 2020, \apjl, 896, L22

\bibitem[{{Mousis} {et~al.}(2019){Mousis}, {Ronnet}, \&
  {Lunine}}]{2019ApJ...875....9M}
{Mousis}, O., {Ronnet}, T., \& {Lunine}, J.~I. 2019, \apj, 875, 9

\bibitem[{{Nakajima} {et~al.}(1992){Nakajima}, {Hayashi}, \&
  {Abe}}]{1992JAtS...49.2256N}
{Nakajima}, S., {Hayashi}, Y.-Y., \& {Abe}, Y. 1992, Journal of Atmospheric
  Sciences, 49, 2256

\bibitem[{{Noack} {et~al.}(2016){Noack}, {H{\"o}ning}, {Rivoldini},
  {Heistracher}, {Zimov}, {Journaux}, {Lammer}, {Van Hoolst}, \&
  {Bredeh{\"o}ft}}]{2016Icar..277..215N}
{Noack}, L., {H{\"o}ning}, D., {Rivoldini}, A., {et~al.} 2016, \icarus, 277,
  215

\bibitem[{{Noack} {et~al.}(2017){Noack}, {Rivoldini}, \& {Van
  Hoolst}}]{2017PEPI..269...40N}
{Noack}, L., {Rivoldini}, A., \& {Van Hoolst}, T. 2017, Physics of the Earth
  and Planetary Interiors, 269, 40

\bibitem[{{Ortenzi} {et~al.}(2020){Ortenzi}, {Noack}, {Sohl}, {Guimond},
  {Grenfell}, {Dorn}, {Schmidt}, {Vulpius}, {Katyal}, {Kitzmann}, \&
  {Rauer}}]{2020NatSR..1010907O}
{Ortenzi}, G., {Noack}, L., {Sohl}, F., {et~al.} 2020, Scientific Reports, 10,
  10907

\bibitem[{{Papaloizou} {et~al.}(2018){Papaloizou}, {Szuszkiewicz}, \&
  {Terquem}}]{2018MNRAS.476.5032P}
{Papaloizou}, J.~C.~B., {Szuszkiewicz}, E., \& {Terquem}, C. 2018, \mnras, 476,
  5032

\bibitem[{{Plotnykov} \& {Valencia}(2020)}]{2020MNRAS.499..932P}
{Plotnykov}, M. \& {Valencia}, D. 2020, \mnras, 499, 932

\bibitem[{{Pluriel} {et~al.}(2019){Pluriel}, {Marcq}, \&
  {Turbet}}]{2019Icar..317..583P}
{Pluriel}, W., {Marcq}, E., \& {Turbet}, M. 2019, \icarus, 317, 583

\bibitem[{{Raymond} {et~al.}(2018){Raymond}, {Boulet}, {Izidoro}, {Esteves}, \&
  {Bitsch}}]{2018MNRAS.479L..81R}
{Raymond}, S.~N., {Boulet}, T., {Izidoro}, A., {Esteves}, L., \& {Bitsch}, B.
  2018, \mnras, 479, L81

\bibitem[{{Ricker} {et~al.}(2015){Ricker}, {Winn}, {Vanderspek}, {Latham},
  {Bakos}, {Bean}, {Berta-Thompson}, {Brown}, {Buchhave}, {Butler}, {Butler},
  {Chaplin}, {Charbonneau}, {Christensen-Dalsgaard}, {Clampin}, {Deming},
  {Doty}, {De Lee}, {Dressing}, {Dunham}, {Endl}, {Fressin}, {Ge}, {Henning},
  {Holman}, {Howard}, {Ida}, {Jenkins}, {Jernigan}, {Johnson}, {Kaltenegger},
  {Kawai}, {Kjeldsen}, {Laughlin}, {Levine}, {Lin}, {Lissauer}, {MacQueen},
  {Marcy}, {McCullough}, {Morton}, {Narita}, {Paegert}, {Palle}, {Pepe},
  {Pepper}, {Quirrenbach}, {Rinehart}, {Sasselov}, {Sato}, {Seager},
  {Sozzetti}, {Stassun}, {Sullivan}, {Szentgyorgyi}, {Torres}, {Udry}, \&
  {Villasenor}}]{2015JATIS...1a4003R}
{Ricker}, G.~R., {Winn}, J.~N., {Vanderspek}, R., {et~al.} 2015, Journal of
  Astronomical Telescopes, Instruments, and Systems, 1, 014003

\bibitem[{{Ronnet} {et~al.}(2017){Ronnet}, {Mousis}, \&
  {Vernazza}}]{2017ApJ...845...92R}
{Ronnet}, T., {Mousis}, O., \& {Vernazza}, P. 2017, \apj, 845, 92

\bibitem[{{Schoonenberg} {et~al.}(2019){Schoonenberg}, {Liu}, {Ormel}, \&
  {Dorn}}]{2019A&A...627A.149S}
{Schoonenberg}, D., {Liu}, B., {Ormel}, C.~W., \& {Dorn}, C. 2019, \aap, 627,
  A149

\bibitem[{{Seager} {et~al.}(2007){Seager}, {Kuchner}, {Hier-Majumder}, \&
  {Militzer}}]{2007ApJ...669.1279S}
{Seager}, S., {Kuchner}, M., {Hier-Majumder}, C.~A., \& {Militzer}, B. 2007,
  \apj, 669, 1279

\bibitem[{{Shaw}(1986)}]{1986JChPh..84.5862S}
{Shaw}, G.~H. 1986, \jcp, 84, 5862

\bibitem[{{Sotin} {et~al.}(2007){Sotin}, {Grasset}, \&
  {Mocquet}}]{2007Icar..191..337S}
{Sotin}, C., {Grasset}, O., \& {Mocquet}, A. 2007, \icarus, 191, 337

\bibitem[{{Stevenson} \& {Lunine}(1988)}]{1988Icar...75..146S}
{Stevenson}, D.~J. \& {Lunine}, J.~I. 1988, \icarus, 75, 146

\bibitem[{{Tchijov} {et~al.}(2004){Tchijov}, {Ayala}, {Leon}, \&
  {Nagornov}}]{2004JPCS...65.1277T}
{Tchijov}, V., {Ayala}, R.~B., {Leon}, G.~C., \& {Nagornov}, O. 2004, Journal
  of Physics and Chemistry of Solids, 65, 1277

\bibitem[{{Tulk} {et~al.}(1997){Tulk}, {Gagnon}, {Kiefte}, \&
  {Clouter}}]{1997JChPh.10710684T}
{Tulk}, C.~A., {Gagnon}, R.~E., {Kiefte}, H., \& {Clouter}, M.~J. 1997, \jcp,
  107, 10684

\bibitem[{{Turbet} {et~al.}(2020{\natexlab{a}}){Turbet}, {Bolmont}, {Bourrier},
  {Demory}, {Leconte}, {Owen}, \& {Wolf}}]{2020SSRv..216..100T}
{Turbet}, M., {Bolmont}, E., {Bourrier}, V., {et~al.} 2020{\natexlab{a}}, \ssr,
  216, 100

\bibitem[{{Turbet} {et~al.}(2020{\natexlab{b}}){Turbet}, {Bolmont},
  {Ehrenreich}, {Gratier}, {Leconte}, {Selsis}, {Hara}, \&
  {Lovis}}]{2020A&A...638A..41T}
{Turbet}, M., {Bolmont}, E., {Ehrenreich}, D., {et~al.} 2020{\natexlab{b}},
  \aap, 638, A41

\bibitem[{{Turbet} {et~al.}(2018){Turbet}, {Bolmont}, {Leconte}, {Forget},
  {Selsis}, {Tobie}, {Caldas}, {Naar}, \& {Gillon}}]{2018A&A...612A..86T}
{Turbet}, M., {Bolmont}, E., {Leconte}, J., {et~al.} 2018, \aap, 612, A86

\bibitem[{{Turbet} {et~al.}(2019){Turbet}, {Ehrenreich}, {Lovis}, {Bolmont}, \&
  {Fauchez}}]{2019A&A...628A..12T}
{Turbet}, M., {Ehrenreich}, D., {Lovis}, C., {Bolmont}, E., \& {Fauchez}, T.
  2019, \aap, 628, A12

\bibitem[{{Unterborn} {et~al.}(2018){Unterborn}, {Desch}, {Hinkel}, \&
  {Lorenzo}}]{2018NatAs...2..297U}
{Unterborn}, C.~T., {Desch}, S.~J., {Hinkel}, N.~R., \& {Lorenzo}, A. 2018,
  Nature Astronomy, 2, 297

\bibitem[{{Vazan} {et~al.}(2013){Vazan}, {Kovetz}, {Podolak}, \&
  {Helled}}]{2013MNRAS.434.3283V}
{Vazan}, A., {Kovetz}, A., {Podolak}, M., \& {Helled}, R. 2013, \mnras, 434,
  3283

\bibitem[{Wagner \& Pru{\ss}(2002)}]{doi:10.1063/1.1461829}
Wagner, W. \& Pru{\ss}, A. 2002, Journal of Physical and Chemical Reference
  Data, 31, 387

\bibitem[{{Wang} {et~al.}(2017){Wang}, {Wu}, {Barclay}, \&
  {Laughlin}}]{2017arXiv170404290W}
{Wang}, S., {Wu}, D.-H., {Barclay}, T., \& {Laughlin}, G.~P. 2017, arXiv
  e-prints, arXiv:1704.04290

\bibitem[{{Zeng} {et~al.}(2019){Zeng}, {Jacobsen}, {Sasselov}, {Petaev},
  {Vanderburg}, {Lopez-Morales}, {Perez-Mercader}, {Mattsson}, {Li}, {Heising},
  {Bonomo}, {Damasso}, {Berger}, {Cao}, {Levi}, \&
  {Wordsworth}}]{2019PNAS..116.9723Z}
{Zeng}, L., {Jacobsen}, S.~B., {Sasselov}, D.~D., {et~al.} 2019, Proceedings of
  the National Academy of Science, 116, 9723

\end{thebibliography}

\end{document}